# Revealing Flare Energetics and Dynamics with SDO EVE Solar Extreme Ultraviolet Spectral Irradiance Observations

Thomas N. Woods[1], Phillip C. Chamberlin[1], Andrew Jones[1], James P. Mason[2], Liying Qian[3], Harry P. Warren[4], Don Woodraska[1], Rita Borelli[1], Francis G. Eparvier[1], Gabi Gonzalez[1]

1. Laboratory for Atmospheric and Space Physics, University of Colorado, Boulder, CO
2. Johns Hopkins University, Applied Physics Laboratory, Laurel, MD
3. High Altitude Observatory, National Center for Atmospheric Research, Boulder, CO
4. Naval Research Laboratory, Washington, DC

## Abstract

NASA's Solar Dynamics Observatory (SDO) Extreme-ultraviolet Variability Experiment (EVE) has been making solar full-disk extreme ultraviolet (EUV) spectral measurements since 2010 over the spectral range of 6 nm to 106 nm with 0.1 nm spectral resolution and with 10-60 sec cadence. A primary motivation for EVE's solar EUV irradiance observations is to provide the important energy input for various studies of Earth's upper atmosphere. For example, the solar EUV radiation creates the ionosphere, heats the thermosphere, and drives photochemistry in Earth's upper atmosphere. In addition, EVE's observations have been a treasure trove for solar EUV flare spectra. While EVE measures the full-disk spectra, the flare spectrum is easily determined as the EVE spectrum minus the pre-flare spectrum, as long as only one flare event is happening at a time. These EVE flare observations provide EUV variability with 0.1 nm spectral resolution and have been used to study flare phases (including the discovery of the EUV Late Phase flare class), flare energetics (plasma temperature variations), corona heating (plasma abundance changes that support nano-flare heating mechanism), flare dynamics (downwelling and upwelling plasma flows during flares from Doppler-related wavelength shifts), and coronal mass ejections (CME) energetics (CME mass and velocity derived from coronal dimming in some EUV lines). A brief review of each of these flare research topics using EVE data are presented. With over 10,000 flares detected in the EVE observations, there is still much to study and to learn about solar flare physics using EVE solar EUV spectra.

## Plain-language summary:

NASA's SDO EVE instrument has been making solar extreme-ultraviolet measurements for more than 15 years, and those observations have included thousands of flare observations. These flare measurements have been important to study flare energetics and dynamics during different flare phases and consequently improving understanding of corona heating, coronal mass ejections associated with flares, and impacts into Earth's ionosphere and thermosphere.

## 1. Introduction

The solar extreme ultraviolet (EUV: 10-120 nm) radiation emerges from the chromosphere, transition region, and corona layers of the solar atmosphere, and the solar soft X-ray (SXR: 0.1-10 nm) radiation emerges from the hotter corona. The H and He emissions dominate much of the EUV range, and the corona is notably hot in the 1 MK to 30 MK range to excite highly ionized Fe, Mg, Si, and many other species. The solar EUV and SXR variability, which can be several percent to more than a factor of 10 dependent on wavelength, is manifested by changing magnetic fields altering the solar atmosphere. The magnetic structures can be characterized by closed magnetic field structures, such as bright active regions above sunspots, and open magnetic field structures, such as dark coronal holes that are a primary source for the fast solar wind. The EUV and SXR emissions are significantly brighter from the closed magnetic field structures. Importantly, the magnetic fields begin to confine the ionized plasma in the corona where corona loops, arcades, prominences, and other structures are clearly observed as being highly dynamic on time scales of seconds to hours. The emergence and evolution of the active regions drive the variability of most EUV and SXR emissions from hours to months and over the 11-year solar activity cycle, and the EUV and SXR emissions are most energetic during flares (e.g., Lean, 2011; Chamberlin et al., 2008). The primary energy source for the solar flares is considered to be magnetic field reconnection in the corona that accelerates particles to very high energy that then leads to heating the corona to more than 10 MK and modifying coronal loops and sometimes creating flare ribbons and post-flare arcades (e.g., Shibata & Magara, 2011; Benz, 2017).

The EUV Variability Experiment (EVE) on the Solar Dynamics Observatory (SDO) was primarily developed to study the solar EUV and SXR spectral irradiance to understand better its variability on times scales from seconds to years, how and why the solar EUV irradiance varies, and its impact on Earth's ionosphere and thermosphere (Woods et al., 2012). This latter objective is a critical link between the Sun and Earth as the solar EUV and SXR irradiance is the primary energy input to the Earth's upper atmosphere to create the ionosphere and heat the thermosphere (e.g., Huba et al., 2005; Solomon et al., 2013). The variations of the solar EUV and SXR radiation directly drive much of the variability in Earth's upper atmosphere and have several space weather impacts, such as affecting satellite operations through changing the atmospheric density and thus satellite drag and enhancing the ionosphere density and thus potentially being disruptive for some of our navigation and communication systems (e.g., Schrijver & Siscoe, 2010).

Due to the entirety of the vacuum-ultraviolet (VUV: 0-200 nm) irradiance from the Sun depositing its energy into the solar atmosphere, none of this radiation makes it to the ground and space-based measurements are required. There have been many challenges in having frequent enough space flights with solar irradiance instruments; consequently, there have been several large gaps in time and in wavelength, such as the "EUV Hole" in the 1980-1990s (Donnelly, 1987). Additionally, these space-based instruments degrade over time, leading to larger uncertainties over time. These factors lead to large data gaps in solar VUV observations, including the entirety of time prior to the space age, years-to-decades data gaps between the earlier missions, as well as even seconds to hours data gaps within instrument observations themselves.

SDO EVE has the goal to fill these short-term data gaps with planned 24/7 observations from 0-105 nm and H I Lyman-alpha (121 nm). Prior to the launch of SDO in 2010, the solar EUV spectral irradiance was measured daily by the Solar EUV Experiment (SEE, 1-195 nm, 0.4 nm resolution) aboard the Thermosphere, Ionosphere, Mesosphere, Energetics, and Dynamics (TIMED) satellite (Woods et al., 2005), and a few spectral bands were measured at higher time cadence by the GOES X-Ray Sensor (XRS, 2 bands in 0.05-0.8 nm range) (Hanser & Sellers, 1996) and by the SOHO Solar EUV Monitor (SEM, 2 bands in 1-120 nm range) (Judge et al., 1998). The GOES XRS measurements have been a key monitor of solar flares for more than 50 years to provide space weather operational alerts for flare events and to provide an indication of corona temperature and emission measure changes during flares (e.g., Garcia, 2000; White et al., 2005). The relevance of the solar EUV irradiance measurements for flare physics was not well recognized until after the EVE observations began in May 2010. The solar irradiance measurements, being full-disk observations, cannot provide spatial information about the flare events, but can provide "flare spectra" by the simple technique of subtracting the pre-flare spectra from the irradiance spectra during the flare event. These flare spectra are most insightful into flare physics when there is just one main flare being active at a time, which turns out to be true most of the time.

The SDO EVE measurements of the solar EUV spectral irradiance have proven to be very useful to study flare phases (including the discovery of the EUV Late Phase flare class), flare energetics (plasma temperature variations), corona heating (plasma abundance changes support nano-flare heating mechanism), flare dynamics (downwelling and upwelling plasma flows during flares from Doppler-related wavelength shifts), and coronal mass ejections (CME) energetics (CME mass and velocity derived from coronal dimming in some EUV lines). We also briefly introduce a new EVE data product called the EVE Level 4 Lines data product, which provides line-profile-fit results for intensity, wavelength shift, and line width for 70 emission features. These emission features are from the chromosphere, transition region, and corona, and so Doppler velocity measurements of those lines can reveal important plasma dynamical behavior during a flare's impulsive phase and gradual phase. A brief review of each of these flare research topics are presented in the context of the SDO 15-year mission for the *Solar Physics* Topical Issue titled "*How SDO Has Revolutionized Our Understanding of the Sun"*. This paper is also provided as a summary of our presentation made at the February 2025 SDO Workshop titled *"A Gathering of the Helio-hive!"*.

## 2. EVE Flare Observations

NASA's Solar Dynamics Observatory (SDO) Extreme-ultraviolet Variability Experiment (EVE) has been making solar extreme-ultraviolet (EUV) spectral measurements for more than 15 years. EVE's Multiple EUV Grating Spectrograph (MEGS) from the University of Colorado makes full-disk spectral measurement with 0.1 nm spectral resolution and 10-60 sec cadence (Woods et al., 2012; Hock et al., 2012). The MEGS-A channel measures the solar EUV spectral irradiance between 6 nm and 37 nm and with 10 sec cadence, but a capacitor short in its CCD camera in June 2014 has limited the MEGS-A observations. That is, the MEGS-A valid solar observations are between May 2010 and June 2014. The MEGS-A channel has two entrance slits with different entrance foil filters to isolate the solar spectrum between 6 nm and 20 nm for MEGS-A slit 1 and between 16 nm and 37 nm for MEGS-A slit 2. The MEGS-B channel measures the solar EUV spectral irradiance between 32 nm and 106 nm with 0.1 nm spectral resolution and with 10-60 sec cadence. MEGS-B CCD camera has not had the capacitor short issue, but MEGS-B has had more degradation than originally anticipated for wavelengths longer than 70 nm. So, MEGS-B is operated now with 3-hour daily observations to reduce solar exposure. To optimize MEGS-B flare measurements, EVE's flight software has been updated to autonomously detect the start of a flare and then begin a 3-hour flare campaign for the MEGS-B channel whenever the solar X-ray radiation goes above a M1 flare level. Two other MEGS channels are the MEGS-SAM channel (part of MEGS-A) for solar soft X-ray (SXR) images with a 1-7 nm passband and the MEGS-P channel (part of MEGS-B) for measuring the bright H I Lyman-alpha emission line brightness. The other key EVE channel is the EUV SpectroPhotometer (ESP) from the University of Southern California (Didkovsky et al., 2012). ESP has multiple photometer channels, each having a spectral passband of about 4 nm and measuring at a higher cadence of 4 Hz. The ESP bands are centered near 3.5 nm, 18 nm, 26 nm, 30 nm, and 36 nm. The ESP 1-7 nm band channel uses a quadrant photodiode and thus provides flare location as well as the solar SXR irradiance. The ESP has not had any flight anomalies and continues to operate 24/7 with 4 Hz cadence.

### 2.1. MEGS-B Flare Campaign Automation

The on-board algorithm to detect flare events uses the ESP quad-diode measurement of the SXR irradiance. ESP's band of 1-7 nm is very sensitive to flare events with changes more than a factor of 30 for the larger X-class flares. To help avoid false triggers for flares, the slope of the SXR time series and the magnitude of the SXR irradiance are both used with a lower limit for the slope and magnitude. The magnitude limit is intended for M1 or larger flares. Smaller flares (below M1) and also very slow rising flares (e.g., some filament eruptions) will not trigger a MEGS-B flare campaign. A disadvantage of this flare-detection algorithm is that the very beginning of the flare's impulsive phase and any flare onset effects are not included in the MEGS-B flare campaign. However, there are sometimes multiple flares during the 3-hour flare campaigns (and daily 3-hour observations), and then all phases of those extra flares are usually observed.

### 2.2. ESP Flare Location Capability

The ESP quad-diode measurement can also be used to provide the flare location. However, this is not a direct measurement of the flare itself but is rather a measurement of the center of intensity of the entire solar disk. So, to calculate the location of the flare, the pre-flare signals must first be subtracted from the quad-diode signals before doing the quad-diode position calculation. Also, as this is not an imaging system, if there is more than a single flare occurring simultaneously, all that ESP can report is the center of intensity (COI) of the flares. This is not a real problem if the flares are from a single active region, but if the flares are significantly separated on the solar disk, the COI of the combined system is all that can be reported.

Complicating the flare location precision of the ESP are some further design considerations. The aspect ratio of the entrance slit of ESP is 10:1 so the sensitivity in the dispersion direction is much higher than the non-dispersion direction. To further complicate the positional analysis there is a transmission grating and two foil filters in the zeroth-order path to the ESP quad-diode, all of these have support meshes that cause diffraction and give rise to a very complicated point spread function at the detector. The combination of these ESP quad-diode design makes the flare locations from ESP less accurate than results from other quad-diode solar position sensors (like GOES-16 XRS) and from solar EUV / SXR imagers.

### 2.3. MEGS-SAM Flare Location Capability

The SAM is a SXR pinhole camera providing a pixel scale of about 9.2 arc seconds. This provides direct imaging of bright active regions, hot loop structures, and flares. Flare locations can be directly identified from these images.

The SAM pinhole aperture is made of a pinhole in a beryllium copper sheet and is mounted on a tantalum shield. SXRs are blocked by the beryllium copper and so only go through the pinhole aperture. Hard X-ray are blocked by the tantalum but can go through the beryllium copper. Normally, the solar HXR intensity is too low to be detected by SAM, but during a flare the Sun emits enough hard X-rays that the SAM tantalum shield acts as a second pinhole with a larger aperture area, that causes a second bright spot at the flare location in the SAM image due to the HXRs. The center of this HXR spot can also be used to calculate the position of the flare.

The MEGS flare catalog (2010-2014) provides a listing of flares observed by EVE-MEGS along with useful information about the flare event, such as the GOES XRS flare magnitude and flare location results from ESP quad-diode, SAM solar SXR images, and AIA solar EUV images. This MEGS flare catalog and also the MEGS-B flare campaign list are available on the SDO EVE data web site (https://lasp.colorado.edu/eve/data_access). The Archival Solar flaRes (ASR) flare catalog (Berretti et al., 2025) is useful for the MEGS-B flare catalog (2010-present) to provide flare location as determined mostly by AIA solar EUV images.

### 2.4. Extracting Flare Spectra from EVE Full-disk Irradiance Measurements

EVE measures the full-disk spectra, and so flare enhancements are on top of the emissions from across the full-disk. It has been found that the flare spectrum can be effectively isolated from the emissions from other regions on the solar disk by subtracting a pre-flare spectrum from the EVE spectrum during a flare event. This step is critical if you wish to quantify the total radiated energy output for flares alone in these irradiance measurements (Chamberlin et al., 2012). This technique is most accurate when only one flare event is happening at a time. For example, Mason et al. (2014) has shown very good comparisons between EVE coronal dimming events for the cooler coronal lines as compared to analyses of the coronal dimming locations in SDO AIA solar EUV images at the same wavelength. Subtracting off the pre-flare spectrum is not required to detect flare events in the EVE spectra, but many of the flare changes in the EUV are sometimes only about 1% change. Thus, the subtraction of the pre-flare spectrum helps to better clarify the flare-only enhancements in the solar EUV spectral irradiance measurements. Furthermore, the subtraction of the pre-flare spectrum is especially important for MEGS-A spectra, which have spectral shifts based on active region locations on the solar disk (Chamberlin, 2016; Gonzalez et al., 2024).

For almost all of the following discussions about EVE flare observations, the pre-flare spectrum has first been subtracted to isolate the "flare spectrum". One simple technique to identify the time for the pre-flare spectrum is to examine the SXR signal from ESP quad-diode (1-7 nm band) or from GOES XRS signal (0.1-0.8 nm) going backwards in time from the SXR peak to find a local minimum between this flare and the previous flare. If there wasn't a previous flare for more than six hours, then one could take the time of the minimum irradiance within six hours before the flare peak.

## 3. EVE Data Products

The EVE data products that are commonly used for flare research are briefly described here. For details of the EVE data processing algorithms and data products, see Woods et al. (2012) and the EVE Calibration and Measurement Algorithms Document (CMAD, Woods et al., 2025a).The EVE Level 0a and 0b data products contain EVE data in engineering units but not in irradiance units, so they are generally not useful for flare research. However, the EVE Level 0CS is a near-realtime Space Weather product that has irradiance measurements for a few lines and bands, so this product can be useful for realtime flare monitoring. The EVE Level 1 and Level 2 data products contain the EVE solar data in irradiance units and with the original observation cadence (e.g., 10-sec for MEGS spectra), so those products are most useful for flare research. The EVE Level 3 data product has daily-averaged data, which are not applicable for flare studies. The EVE Level 4 Spectral Model and Level 4 Lines products have 1-min cadence results, so they can also be useful for flare research.

### 3.1. EVE Level 0CS Space Weather Product

The EVE Level 0CS Space Weather product has very low latency containing one-minute averages of up to 6 integration sets and that are reported within about 10 seconds after the end of the UTC minute. This low-latency space weather product is made using the SDO S-band packet stream. Because of the lower data rates for the S-band packets, only one out of every 40 integrations from the set of all photometers (ESP and MEGS-P) are available in each 10-second cadence EVE S-band packet. The complete set of photometer data is available later in Level 1 processing.

The EVE L0CS data are stored as fixed-width-column ASCII-text files and with a file per day. Each line in the L0CS file includes the ESP photometer irradiances, MEGS-P Lyman-alpha broadband irradiance, ESP quadrant diode currents (useful for flare location calculation), ESP dark diode current (useful for GEO energetic particle flux), proxy model of SOHO SEM 30nm-channel irradiance, and proxy model of XRS-B 0.1-0.8 nm irradiance based on ESP quadrant diode measurement (Hock et al., 2013).

A special low-latency 10-second cadence diode irradiance product is also available that uses the EVE L0CS data product. This special product has lower latency (about 12 sec) than that possible from the operational GOES XRS data products (which have about 60-sec latency). This special product was initially developed to support the NASA sounding rocket flare campaign in April 2024 from the NASA Poker Flats rocket range.

### 3.2. EVE Level 1 and Level 2 Data Products

EVE Level 1 processing algorithms convert the EVE L0b solar data into irradiance units and retain the native instrument cadence with reversible (dark subtraction, integration rate) and irreversible corrections (particle filtering, integration from images to spectra) including radiometric calibration (gain, responsivity) based on pre-launch ground calibration results [see Hock et al. (2012) for MEGS calibration results and Didkovsky et al. (2012) for ESP calibration results]. Level 1 processing for MEGS-A and MEGS-B produces radiometrically calibrated spectral irradiances that have reduced the Level 0b 2d images into integrated fixed-wavelength spectra. Two spectra are produced in each MEGS-A integration for slit 1 and slit 2 from each integration. The ESP Level 1 and MEGS-P Level 1 products also have degradation applied and retain the native instrument 4 Hz integration rates, so those products are most useful for flare studies in the 4 Hz to 0.1 Hz time range.

Level 2 EVS processing merges the Level 1 MEGS-A and MEGS-B full-disk solar EUV spectra together into one spectrum at 0.02 nm sampling, retaining 0.1 nm instrument resolution at a 10-sec cadence. The Level 2 spectra contain the long-term degradation corrections and adjustments from the rocket EVE underflight measurements. Level 2 data at 10-sec cadence are only available prior to 2018 when the MEGS-B integration rate was increased to 60 seconds to improve upon the MEGS-B measurement precision. The Level 2b EVS product contains the full-disk solar spectra over the entire mission to include those 60-sec MEGS-B spectra after 2018 and with the MEGS-A and -B 10-sec spectra prior to 2018 with an average over one-minute. The Level 2 EVS product is commonly used to study flare spectra at 10-sec and 60-sec cadences.

The Level 2 EVL products contain integrated solar emission line irradiances from the MEGS spectra, integrated broadband irradiances from the MEGS spectra, broadband diode irradiances from ESP and MEGS-P averaged to the spectral cadence, and the fractional ESP quad diode signals that can be useful for solar position information. The spectral lines in the Level 2 EVL product span a wide temperature range from the chromosphere through the corona, and thus they are useful to study flare physics in different layers of the solar atmosphere.

Interested users of the EVE data may wish to take advantage of the LASP Interactive Solar Irradiance Data Center (LISIRD). This interactive data center is useful for identifying data of interest before downloading large quantities of data.

### 3.3. EVE Level 4 Spectral Model Product

Because of the spectral gaps in the EVE solar EUV spectral irradiance data products, there is an EVE Level 4 Spectral Model product that uses ESP 1-7 nm diode irradiance as a variability proxy for this model based on CHIANTI model spectra (Dere et al., 1997; Landi et al., 2013; Del Zanna et al., 2015). The spectral gaps to fill include 0.1 nm to 6 nm for the full SDO mission and the 6 nm to 33 nm range after June 2014 when MEGS-A CCD camera had its capacitor failure. The Level 4 Spectral Model is also useful to fill time gaps between the MEGS-B observations (33 nm to 106 nm range). The original development of this Level 4 model product (Woods et al., 2008; Woods and Elliott, 2022) was done for the Solar Radiation and Climate Experiment (SORCE) X-ray Photometer System (XPS) and Thermosphere, Ionosphere, Mesosphere, Energetics, and Dynamics (TIMED) Solar EUV Experiment (SEE). This SORCE-XPS / SEE-XPS Level 4 Spectral Model is modified slightly for EVE Level 4 Spectral Model by using the ESP 1-7 nm irradiance instead of the XPS 1-7 nm irradiance for the variability proxy. The EVE Level 4 Spectral Model covers the 0.01 nm to 105.99 nm range with 0.02 nm bins and time cadence of 1 min. The EVE Level 4 Spectral Model product is useful when MEGS-A or MEGS-B spectra are not available during a flare event.

### 3.4. EVE Level 4 Lines Fit Product

A new EVE data product was developed in 2024-2025 to provide wavelength shifts of 70 different solar EUV lines observed by EVE MEGS. This product, called the EVE Level 4 Lines data product, can be useful to study Doppler velocity of flaring plasma after corrections are made to account for MEGS instrumental wavelength shifts sensitive to the flare location on the solar disk (e.g., Chamberlin, 2016). The EVE Level 4 Lines algorithm includes fitting a Gaussian to the spectral feature of interest plus two additional Gaussians for adjacent lines (blends) in the wings of the feature of interest along with a linear background fit of the solar continuum. The Gaussian fit for the feature of interest provides fitted results for the wavelength center, spectral width, and intensity. The two wing Gaussian fits only allow adjustment of the intensity for those adjacent lines; that is, the wavelength centers and line widths are fixed for the wing lines. In addition, the EVE Level 4 Lines algorithm does additional data analysis for flare events detected in the GOES XRS time series. For those detected flares, a pre-flare MEGS spectrum is subtracted from the MEGS spectra during the GOES flare period to isolate the "flare spectra", and then the same spectral feature fitting algorithm (three Gaussians and linear background) is fit to the flare spectra. The FLARE_FLAG in the EVE Level 4 Lines product indicates when the additional fits are done for the "flare spectra". The selection process for the pre-flare spectrum is an algorithm developed with EVE spectra to study coronal dimming events (Mason et al., 2019), and this algorithm involves examining the irradiance variability over a six-hour period to find the minimum irradiance in between adjacent flares. This autonomous selection contributes some small uncertainties of a few percent to the flare spectra irradiance level, more so for the cooler emission lines that do not have strong flare variability, but it doesn't usually impact the wavelength shifts for the flare lines.

There are many different spectral lines in the EUV range that are emitted primarily from the transition region and corona. Many of these emission lines are blended for MEGS spectral resolution of 0.1 nm, so we first analyzed all of the hundreds of the emission features in the MEGS-A and MEGS-B spectra to identify the best emission features to include in the EVE Level 4 Lines product. The spectral features in the EVE Level 4 Lines product were selected based on being unblended (or limited blends) or being a bright feature that is desired for research by the solar physics and terrestrial upper atmosphere communities. The process of selecting the 70 spectral features in the EVE Level 4 Lines Fit product will be provided in a separate paper about this data product. For here, it is important to know that about half of those lines are from the transition region and about half are from the corona. There are also a few lines with contributions from the hot corona (> 6 MK), which show up predominantly during flare events.

It also important to know that here are 42 spectral features (9 for MEGS-A and 33 for MEGS-B) that have their line blend error estimate below 30 km/s, and those spectral features are considered the best for studying the flare dynamics, as discussed later in Section 6. This list of the best 42 spectral features for flare studies is provided in Table 1. The full line list, along with discussions about the line blend error estimates and corrections for MEGS instrumental wavelength shifts, will be provided in the EVE Level 4 Lines Fit product paper.

**Table 1**. List of 42 Spectral Features in EVE Level 4 Lines Data Product Most Useful for Flares

| 1 | 2 | 3 | 4 | 5 | 6 | 7 | 8 | 9 | 10 |
|---|---|---|---|---|---|---|---|---|---|
| Index | Reference λ (nm) | QS DEM λ (nm) | AR DEM λ (nm) | FL DEM λ (nm) | Blend Error (km/sec) | Temp MK Mean | Temp MK Range | Primary Element (purity %) | Dominant Ions |
| 1 | 9.393 | 9.393 | 9.393 | 9.393 | 16 | 8.00 | 4.00-12.00 | Fe (98%) | Fe XVIII |
| 5 | 13.113 | 13.112 | 13.111 | 13.111 | 27 | 0.63 | 0.40- 1.00 | Fe (95%) | Fe VIII |
| 9 | 17.107 | 17.107 | 17.107 | 17.109 | 22 | 0.63 | 0.40- 1.50 | Fe (98%) | Fe IX |
| 14 | 20.176 | 20.204 | 20.204 | 20.203 | 9 | 1.50 | 1.00- 1.50 | Fe (97%) | Fe XIII |
| 16 | 21.137 | 21.133 | 21.133 | 21.133 | 4 | 1.50 | 1.50- 2.50 | Fe (96%) | Fe XIV |
| 21 | 24.919 | 24.920 | 24.921 | 24.919 | 12 | 2.50 | 2.50- 8.00 | Ni (98%) | Ni XVII |
| 24 | 27.039 | 27.047 | 27.051 | 27.047 | 27 | 0.40 | 0.40- 0.63 | Mg (96%) | Mg VI, Fe XIV |
| 26 | 28.416 | 28.416 | 28.416 | 28.415 | 4 | 2.50 | 1.50- 8.00 | Fe (98%) | Fe XV |
| 28 | 33.540 | 33.538 | 33.541 | 33.542 | 24 | 2.50 | 2.50-10.00 | Fe (100%) | Fe XVI / Fe XXI, Mg VIII |
| 30 | 36.076 | 36.079 | 36.077 | 36.076 | 14 | 2.50 | 2.50-12.00 | Fe (98%) | Fe XVI |
| 31 | 36.812 | 36.804 | 36.806 | 36.804 | 10 | 1.50 | 1.50- 2.50 | Mg (31%) | Mg VII / Mg IX, Fe XIII |
| 34 | 44.570 | 44.570 | 44.570 | 44.570 | 3 | 4.00 | 4.00-20.00 | S (100%) | S XIV |
| 35 | 46.525 | 46.523 | 46.523 | 46.523 | 1 | 0.63 | 0.40- 0.63 | Ne (98%) | Ne VII |
| 36 | 46.985 | 46.985 | 46.984 | 46.985 | 3 | 0.15 | 0.10- 0.40 | Ne (98%) | Ne IV |
| 37 | 49.941 | 49.941 | 49.941 | 49.941 | 0 | 2.50 | 1.50-20.00 | Si (100%) | Si XII |
| 38 | 50.800 | 50.795 | 50.795 | 50.795 | 1 | 0.10 | 0.04- 0.25 | O (98%) | O III |
| 39 | 52.066 | 52.066 | 52.066 | 52.067 | 1 | 2.50 | 1.50-20.00 | Si (99%) | Si XII |
| 40 | 52.579 | 52.579 | 52.579 | 52.579 | 1 | 0.10 | 0.06- 0.25 | O (100%) | O III |
| 41 | 53.824 | 53.823 | 53.825 | 53.822 | 8 | 0.10 | 0.08- 0.15 | C (88%) | C III |
| 42 | 54.199 | 54.211 | 54.211 | 54.209 | 9 | 0.15 | 0.10- 0.40 | Ne (99%) | Ne IV, Ca IX, Fe XII |
| 43 | 54.389 | 54.390 | 54.397 | 54.391 | 23 | 0.15 | 0.08- 0.25 | Ne (93%) | Ne IV, Ca VII, Fe XIV |
| 44 | 55.002 | 55.004 | 55.003 | 55.004 | 2 | 1.50 | 1.00-12.00 | Al (99%) | Al XI |
| 45 | 55.445 | 55.443 | 55.443 | 55.443 | 1 | 0.15 | 0.08- 0.25 | O (98%) | O IV |
| 47 | 57.230 | 57.230 | 57.230 | 57.230 | 2 | 0.25 | 0.15- 0.63 | Ne (99%) | Ne V |
| 49 | 58.433 | 58.434 | 58.434 | 58.433 | 1 | 0.04 | 0.02- 0.08 | He (99%) | He I |
| 51 | 59.959 | 59.959 | 59.959 | 59.959 | 1 | 0.10 | 0.06- 0.25 | O (100%) | O III |
| 52 | 60.982 | 60.981 | 60.980 | 60.982 | 9 | 0.15 | 0.08-20.00 | O (61%) | O IV / O III, Mg X, Ni XXIV |
| 53 | 62.493 | 62.494 | 62.494 | 62.496 | 7 | 1.00 | 1.00-12.00 | Mg (98%) | Mg X |
| 54 | 62.973 | 62.973 | 62.973 | 62.973 | 0 | 0.25 | 0.10- 1.50 | O (95%) | O V |
| 56 | 76.515 | 76.513 | 76.521 | 76.513 | 22 | 0.15 | 0.10- 0.40 | N (98%) | N IV, Fe VII |
| 57 | 77.043 | 77.041 | 77.043 | 77.041 | 4 | 0.63 | 0.40- 8.00 | Ne (99%) | Ne VIII |
| 58 | 78.769 | 78.772 | 78.775 | 78.773 | 7 | 0.15 | 0.06- 0.25 | O (98%) | O IV, S X |
| 59 | 79.019 | 79.020 | 79.020 | 79.020 | 0 | 0.15 | 0.08- 0.40 | O (99%) | O IV |
| 61 | 90.409 | 90.409 | 90.414 | 90.409 | 11 | 0.04 | 0.02- 0.10 | C (100%) | C II |
| 62 | 92.320 | 92.322 | 92.319 | 92.322 | 6 | 0.15 | 0.06- 0.25 | N (89%) | N IV, Fe III |
| 63 | 93.338 | 93.338 | 93.334 | 93.338 | 8 | 0.25 | 0.10- 1.00 | S (99%) | S VI |
| 64 | 94.974 | 94.974 | 94.974 | 94.974 | 0 | 0.02 | 0.01- 0.40 | H (93%) | H I, Si IX |
| 65 | 97.254 | 97.254 | 97.254 | 97.254 | 0 | 0.02 | 0.01- 2.50 | H (99%) | H I |
| 66 | 97.702 | 97.702 | 97.702 | 97.702 | 0 | 0.08 | 0.02- 0.25 | C (98%) | C III |
| 67 | 102.572 | 102.572 | 102.572 | 102.572 | 0 | 0.02 | 0.01- 4.00 | H (99%) | H I |
| 68 | 103.191 | 103.191 | 103.191 | 103.191 | 0 | 0.25 | 0.15- 6.30 | O (100%) | O VI |
| 69 | 103.689 | 103.703 | 103.693 | 103.702 | 19 | 0.04 | 0.02- 0.10 | C (100%) | C II |

## 4. Flare Phases

### 4.1. Standard Flare Phases: Impulsive Phase and Gradual Phase

The two main flare phases are the impulsive phase and the gradual phase. In the X-ray range, many flares have intense hard X-ray emissions during the impulsive phase followed by intense soft X-ray emissions during the gradual phase. Neupert (1968) noted that the integration of the hard X-ray intensity over time is a good indicator for the soft X-ray profile during the gradual phase. Without routine hard X-ray flare observations, the inverse Neupert effect can be employed by taking the derivative of the GOES XRS SXR time series to estimate the hard X-ray profile during an assumed impulsive phase. By the nature of identifying and classifying flares based on the XRS-B (0.1-0.8 nm) irradiance magnitude, all identified X-ray flares have a gradual phase, but not all flares have a significant impulsive phase for the EUV emission features. In general, the compact (smaller) flares often don't have a clear indication of an impulsive phase; whereas, the eruptive (larger) flares usually have an impulsive phase and often also a coronal mass ejection (CME) associated with the eruptive flare (e.g., Hock, 2012; Hudson, 2011; Gopalswamy et al., 2010; Priest & Forbes, 2002).

For the EVE solar EUV spectral measurements during the flares, the chromospheric and transition lines are usually more intense during the impulsive phase and with a dimmer gradual phase component, and the hotter coronal lines are usually more intense during the gradual phase and have little or no impulsive phase component (e.g., Woods et al., 2011). For our discussions here, we refer to cool corona emissions for those with temperatures of 0.3-2 MK, warm corona for 2-8 MK, hot corona for 8-20 MK, and super-hot corona for above 20 MK. There are 100s of emission features in the EVE spectra, so the timing of when each emission reaches its peak during the gradual phase follows a typical progression from the hotter emissions peaking soon after the impulsive phase followed by the cooler emissions peaking several seconds, even minutes, later (e.g., Chamberlin et al., 2012; Thiemann et al., 2017). These behaviors of EVE emission lines during the impulsive and gradual phases confirm the standard flare model (CSHKP: Carmichael, Sturrock, Hirayama, Kopp, and Pneuman, e.g. Priest & Forbes, 2002). This flare model describes the generation of energetic particles by magnetic reconnection in the corona and then deposition of many of those particles into the chromosphere during the early part of the impulsive phase, followed by chromospheric evaporation of hot plasma into the coronal loops to raise the loop's plasma temperature to well above 10 MK during the later part of the impulsive phase, then rising (expansion) of these heated loops (and sometimes flare ribbons), which then cool and relax back during the gradual phase. In general, the majority of the flare energy is released during the impulsive phase, and the resulting flare evolution in the corona occurs during the gradual phase. Some review papers with more details about the solar multi-wavelength observations during the different flare phases include Benz (2017), Fletcher et al. (2011), and Hudson (2011).

### 4.2. Additional Flare Phases: Onset, Coronal Dimming, and EUV Late Phase

Examples of the flare phases with EVE measurements are shown for two flare events in Figure 1. The first flare example is the X2.2 flare on 15 February 2011, and the second one is the X6.9 flare on 9 August 2011. Those two flares are also used as examples in the other upcoming sections. The top panels show the GOES XRS 0.1-0.8 nm irradiance times series, and the bottom panels show four different emission features that exemplify the different flare phases. Those emission features have its pre-flare irradiance first subtracted and then normalized to its peak irradiance. The transition region emissions, like the He II 30.4 nm, are the first emission features to increase during the impulsive phase (red), which is specified here as starting when the He II emission begins to increase and until the peak of the XRS irradiance. The gradual phase (blue) is defined here as the time period between the 50% points of the XRS irradiance peak (vertical dashed line). The hot coronal emissions in the SXR (GOES XRS) and Fe XX 13.3 nm have strong gradual phase peaks but little impulsive phase contribution. The other cooler emissions peak after the SXR gradual phase peak as the heated post-flare loops cool (e.g., Thiemann et al., 2017).

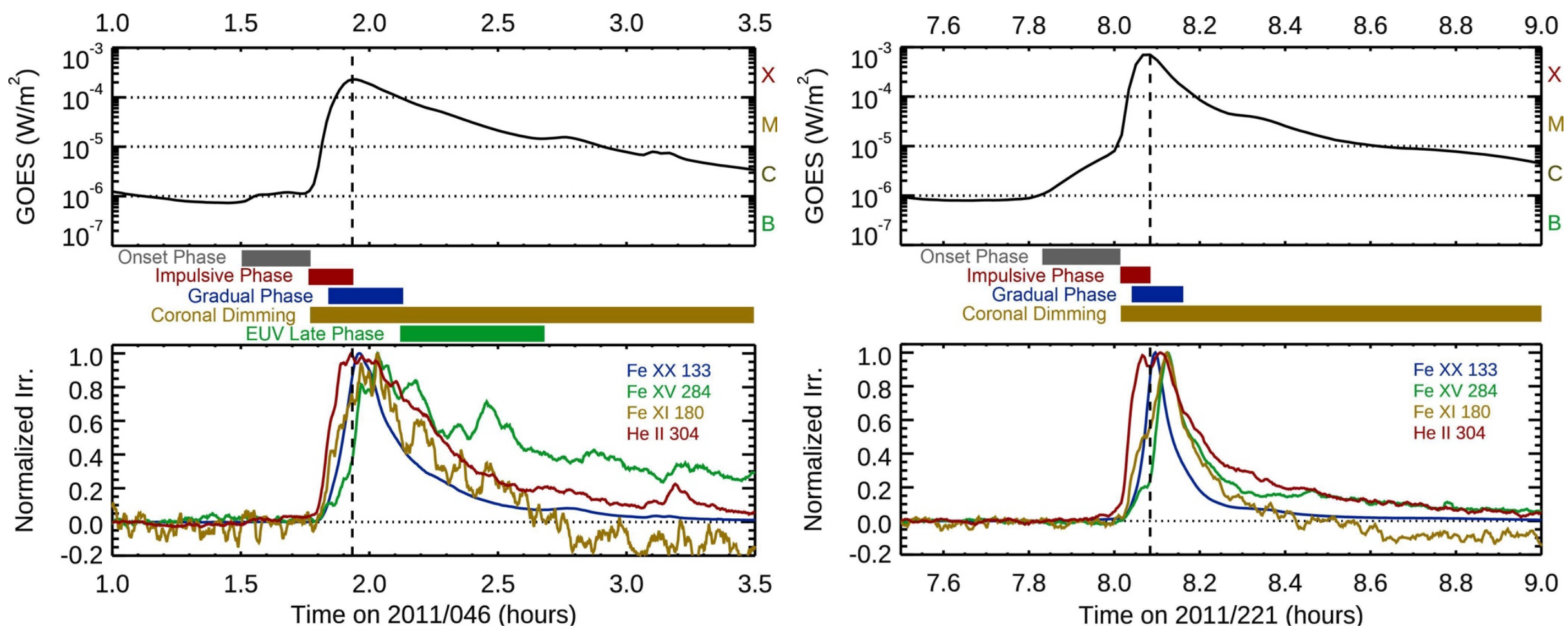

**Figure 1**. Examples of the flare phases observed in the EVE spectra are shown for the X2.2 flare on 15 February 2011 (left plots) and for the X6.9 flare on 9 August 2011 (right plots). These two flares have all of the flare phases discussed here, but many flares, especially the smaller flares, do not have all five flare phases.

In addition to the impulsive and gradual phase components, the EVE flare observations also reveal other phases of some flares that were not initially expected at the start of the SDO mission to be seen in full-disk irradiance spectra. One is that the cooler coronal emissions, such as Fe XII at 19.2 nm, have dimming of a few percent during eruptive flares with CMEs. While coronal dimming has been observed frequently in the regions around a flare in solar EUV images (as early as the 1990s, Thompson et al., 1998), it had not been observed in the solar EUV irradiance spectra prior to the SDO mission. From the EVE observations, the cooler coronal emissions, such as from Fe IX through Fe XII, can have weak impulsive phase contribution and usually larger gradual phase contribution and sometimes a decreased (negative) irradiance relative to the pre-flare level during the coronal dimming period. See Section 7 for more details about some of the coronal dimming results with the EVE spectra and how this result can be extended to predicting CMEs on stars.

Another interesting flare type highlighted by the EVE flare spectra is the EUV Late Phase flare. The EUV Late Phase flare is characterized by having the usual impulsive and gradual phases followed by large secondary peaks in the EUV emissions at moderate plasma temperature of a few MK (e.g., Fe XVI at 33.5 nm) but without any (or very little) increase in the SXR (hotter corona) emissions (Woods et al., 2011). There are times when the X-ray has extra flares after the main flare, along with the other coronal emissions also increasing, and those extra flares are not considered part of the EUV Late Phase period. For example, the left panel in Figure 1 shows increased X-ray emission (extra flares) near 2:50 UT and 3:10 UT. The flare phase color-coded boxes in Figure 1 are provided only for the main (first, larger) flare.

These late peaks, thus the name EUV Late Phase, can show up minutes to an hour after the SXR gradual phase peak and are sometimes brighter than the first gradual phase peaks of the same wavelength. Studying the AIA solar EUV images, in particular the Fe XVI at 33.5 nm, has revealed that the EUV Late Phase flares tend to have a quadrupole (or more complex) active region configuration whereby the flare starts in heating a lower coronal loop that rises to interact with an upper coronal loop and then both sets of coronal loops begin to cool and relax after the impulsive phase. Because the loop cooling conduction rate is proportional to loop length (e.g., Klimchuck et al., 2008), the smaller (lower) loop has its gradual phase peak first (normal peak associated with SXR gradual phase peak), and then the larger (higher) loop has its gradual phase peak later (the EUV Late Phase flare second peak). Woods (2014) studied the frequency and timing of the EUV Late Phase flares over a few solar cycles and found that about 5% of C class flares and more than 20% of M and X class flares can be characterized as EUV Late Phase flares. Woods (2014) also reports that it is more common for the EUV Late Phase flares to occur before and after solar cycle minimum when there are a limited number of active regions present at the same time. This distinctive type of flare is also intriguing for its potential to have a prolonged effect on Earth's ionosphere and thermosphere, as discussed later in Section 9.

Another flare phase that can be studied with the EVE spectra is the flare onset phase, which is the small increase in the X-ray irradiance prior to the impulsive phase. The onset phase is intriguing because it could provide a short-term (several minutes) forecast (nowcast) for a flare event. Hudson et al. (2021), Battaglia et al. (2023), Telikicherla et al. (2024), and Hudson (2025) show that the SXR emissions have a sudden temperature increase to above 10 MK without much brightness change during the flare onset phase. Furthermore, Kniezewski et al. (2024) reveal that some EUV emissions can increase 2-3 hours prior to the flare impulsive phase. Examining the EVE solar EUV spectra during the onset phase of their flare events might reveal additional findings about the physics of the flare onset phenomena.

## 5. Flare Energetics and Abundances

A fundamental property of solar flares is their production of extremely high plasma temperatures, often exceeding 10 MK. However, deriving robust characterizations of the plasma temperature distribution during these events has historically proven challenging. Grazing incidence imaging instruments, such as the X-Ray Telescope (XRT) on Hinode (Golub et al., 2007) or the Soft X-ray Telescope (SXT) on Yohkoh (Tsuneta et al., 1991; Acton, 2016), access the highest temperatures effectively but provide limited coverage at lower temperatures, complicating efforts to characterize the full temperature distribution. Conversely, normal incidence, multi-layer telescopes, such as the Transition Region and Coronal Explorer (TRACE, Handy et al., 1999) and the SDO Atmospheric Imaging Assembly (AIA, Lemen et al., 2012), efficiently capture lower temperatures but are constrained at higher temperatures. Similarly, earlier spectrally resolved instruments, such as the Bragg Crystal Spectrometer (BCS) on Yohkoh (Culhane et al., 1991), lacked sufficient wavelength coverage to determine comprehensive temperature distributions. With nearly complete EUV spectral coverage, EVE has provided unique insights into the high-temperature plasma distribution during solar flares.

Warren (2014) developed a method to compute the differential emission measure (DEM) from EVE spectral observations by approximating the volume DEM as a sum of Gaussian functions in log-temperature space. EVE spectra, including emission lines from Fe XV to Fe XXIV (corresponding to 2 to 30 MK) and continuum from 6 to 20 nm, are pre-processed to remove non-flaring emission. The DEM is then used to forward model the observed spectrum. The presence of the continuum over a broad wavelength range is important because it allows for measurements of absolute abundances. The DEM forward model uses a least-squares minimization to determine each DEM Gaussian component's magnitude. Additionally, the method incorporates electron density constraints through the analysis of density sensitive Fe XXI line ratios. To further enhance the accuracy at high temperatures, GOES soft X-ray flux measurements can be included as an additional constraint. This comprehensive analysis revealed that flare plasmas exhibit broad temperature distributions across all flare phases, highlighting significant limitations inherent to the commonly employed isothermal approximation, which fails to account adequately for the plasma's complex thermal structure.

Measurements of temperature distributions have allowed for more precise determinations of plasma composition during solar flares. It is now well established that the composition of solar plasma is organized according to first ionization potential (FIP), implying that abundance variations are determined by processes occurring in the solar chromosphere and transition region, where low-FIP elements are ionized and high-FIP elements are neutral (Laming, 2004). These variations are thus a critical signature of mass and energy transport through the solar atmosphere. However, measuring composition variations can be challenging because they are difficult to differentiate from temperature effects. The DEMs computed from EVE spectra provide a robust means to study abundance variations in flares while properly accounting for temperature effects.

Warren (2014), in their study of EVE observations for 21 flares, reported that plasma composition remained consistently close to photospheric values, with a mean FIP bias of $1.17 \pm 0.22$, where 1.0 corresponds to photospheric abundances. This finding supports the interpretation that evaporated plasma predominantly originates deep within the chromosphere, contrasting with some earlier studies that frequently indicated enhanced low-FIP element abundances in flare plasma (e.g., Fludra & Schmelz, 1999). Examples of two events are shown in Figure 2. Like many events from this study, the FIP measurements show some variability during the impulsive phase. The calculation of the DEM is most challenging during the impulsive phase when the temperatures are highest, so it is not clear if this is an actual effect or due to systematic issues with the analysis. To partially address this issue, deriving DEM profiles during flares using multiple instruments (multi-wavelengths) can provide a wider coverage of plasma temperature. For example, Caspi et al. (2014) used EVE and RHESSI spectra to cover a wider temperature range and found bimodal-temperature DEM profiles that are similar to that found by Warren (2014) and shown in Figure 2 during the flare's impulsive phase. Additional spectral diagnostics results of the flaring plasma density, temperature, and abundance variations using EVE spectra are provided by Milligan et al. (2012), Petkaki et al. (2012), Del Zanna and Woods (2013), Warren et al. (2013), Milligan (2014), and Warren et al. (2016). Furthermore, Aschwanden et al. (2015) found that combining EVE spectral analysis with GOES XRS or RHESSI data provides even more accurate DEM results.

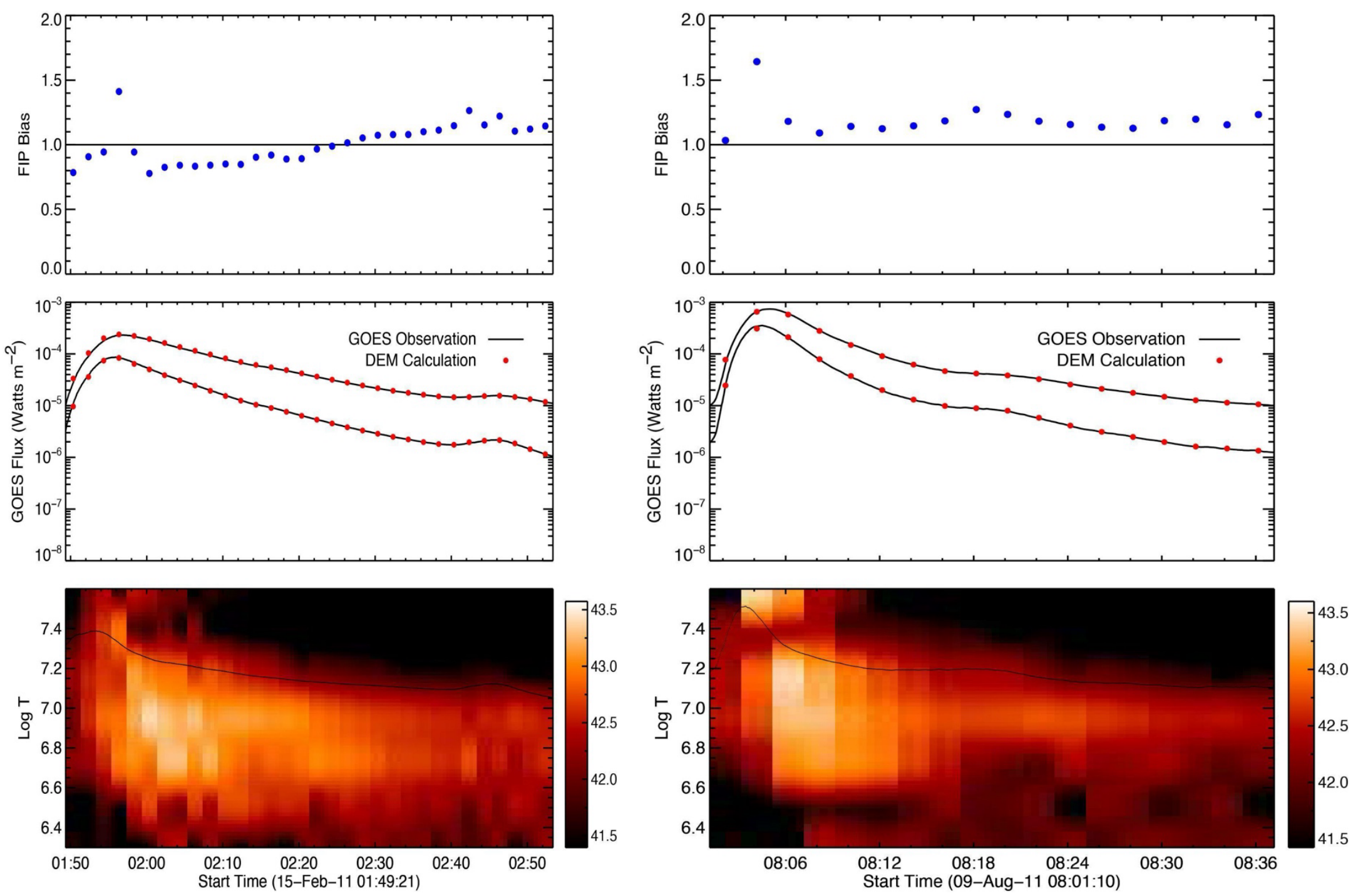

**Figure 2**. Temperature and abundance measurements for two events from the Warren et al. (2014) study of abundance variations in flares observed with EVE. The left plots are for the X2.2 flare on 15 February 2011, and right plots are for the X6.9 flare on 9 August 2011. The three vertical panels show the FIP bias, GOES flux, and DEM distribution as a function of time and temperature. The DEM panel also includes the isothermal temperature derived from GOES as a function of time (black line). This study showed that the temperature distributions in flares are broad and that the composition of flare plasma is generally close to photospheric. There is generally some additional variability during the impulsive phase, when the temperatures are highest.

## 6. Flare Dynamics from Wavelength Shifts

The larger flares, above about M5, have enough flare intensity to provide good measurement precision of the flare variability and associated Doppler wavelength shift in the EVE Level 4 Lines data product. For the smaller flares, the amount of flare variability is only a couple percent for most EUV features, which is comparable to the MEGS measurement precision for individual spectra. The EVE Level 4 Lines product does use 1-min averaged spectra, versus the individual 10-sec spectra, to help improve the measurement precision by about a factor of two. The limited exposure time for MEGS over much of the SDO mission has reduced the number of X-class flares observed by MEGS. In particular, MEGS-A only observed in 2010-2014, and MEGS-B observations are limited to a few hours per day. Furthermore, there was weaker solar activity between 2010 and 2021 for solar cycle 24. Nonetheless, there are 42 X-class flare observations for MEGS-A and 103 X-class flare observations for MEGS-B. The number of coincident X-class flares for both MEGS-A and MEGS-B is just 19.

While we have known about MEGS-A instrument having wavelength shifts related to flare location (Chamberlin, 2012), we have only recently discovered from new raytrace modeling of MEGS-B optical system with active regions at different positions across the solar disk that there are MEGS-B instrumental wavelength shifts when the active regions or flares are near the limb. These recent raytrace analyses indicate wavelength shifts equivalent to as large as +/- 35 km/s Doppler shifts at the out-of-focus wavelengths where the MEGS-B CCD flat plane is most offset from the MEGS-B Rowland circle (in-focus) surface. Thus, much of the "Doppler velocity" results reported by Hudson et al. (2022) and Fitzpatrick & Hudson (2023) for the active region rotation period in August-September 2012 can be explained with MEGS-B instrumental wavelength shifts as function of active region location on the solar disk. These new MEGS-B raytrace model results will be discussed more in a follow-up paper about the EVE Level 4 Lines Fit product. For the discussion here, the example shown for Doppler velocity results during a X-class flare is limited to a flare near disk center where there is little or no additional instrumental wavelength shifts. In other words, it is critical to first correct for the MEGS instrumental wavelength shifts related to active region / flare location before converting the EVE Level 4 Lines wavelength shifts into Doppler velocity estimates for any flares that are not near solar disk center.

The first X-class flare during the SDO mission was the X2.2 flare on 15 February 2011 with GOES XRS flare peak time of 1:56 UT and location of S22-W13 on the solar disk. The following discussion about the Doppler velocity results during that flare does not involve first correcting for the MEGS instrumental wavelength shift, thus an additional uncertainty of about 10 km/s needs to be added to the typical EVE Level 4 Lines wavelength shift fitting uncerainty equivalent to about 5 km/s. From examining the EVE Level 4 Lines data time series during this flare, the time of maximum Doppler wavelength shift from the pre-flare wavelength is during the flare's impulsive phase prior to the XRS flare peak time. A plot of the maximum Doppler velocity for the "low-error" spectral features as a function of line formation temperature is shown in Figure 3. The "low-error" spectral features are those listed in Table 1 with Blend Error less than 30 km/sec. All of the chromosphere, transition region, and cool corona emission features with formation temperature less than 1.0 MK have a maximum Doppler velocity that is positive and thus indicating downflow of flare-affected plasma in those solar layers. The average of these positive Doppler velocities is 77 km/sec +/- 28 km/sec. In contrast, the warmer corona emissions with formation temperature more than 1.0 MK mostly have negative Doppler shifts and thus indicating upflow of flare-affected plasma for an eruptive flare, perhaps even a sign for a coronal mass ejection (CME) event. There isn't as much consistency for the blue (upward) Doppler velocity values, so instead of noting the average, we note the range of Doppler velocities with largest blueshift by the Fe XIV feature (21.14 nm) of - 213 km/sec to a redshift of +50 km/sec for the hotter Fe XVIII (9.39 nm) feature.

This Doppler velocity behavior as function of line formation temperature for this X2.2 flare is similar for other EVE-observed flares that occur near disk center. More importantly, this Doppler velocity behavior with EVE spectra is consistent with flare Doppler velocity behavior established during the impulsive phases of flares with high-spectral-resolution spectroheliograms from SOHO Coronal Diagnostic Spectrometer (e.g., Czaykowska et al., 1999; Brious and Phillips, 2004) and from Hinode EUV Imaging Spectrometer (e.g., Milligan and Dennis, 2009). Brosius and Phillips (2004) relate this Doppler velocity behavior to the chromospheric evaporation process with hot upflowing (about 100 km/s, blueshift) plasma and cool downflowing (about 50 km/s, redshift) plasma happening simultaneously during the flare's impulsive phase. The transition from redshifts to blueshifts between 1 and 2 MK, as seen in the EVE result, is also noted by Milligan and Dennis (2009) for the Hinode EIS observations.

There is concern that there could be line blends for any of these MEGS spectral features as MEGS spectral resolution is only 0.1 nm. While analysis for line blends suggest small Blend Errors in Table 1 for the emission features included in Figure 3, the outliers (redshift instead of blueshift) for the warm coronal lines of Al XI and Fe XVIII might be an indication of larger (unexpected) blend errors for those two lines. Examining the time series of the Doppler wavelength shifts during the flare can help clarify the potential for additional line blends and also reveal the nature of the flare dynamics. Figure 4 shows the time series of six emission features from the transition region and corona during this X2.2 flare on 15 February 2011. The emissions in the left panels of Figure 4 are transition-region emissions that highlight the impulsive phase of the flare. For one, the transition-region emission irradiances (green lines) peak slightly before the XRS gradual phase peak of 1:56 UT. Secondly, the Doppler shifts for these transition-region emissions indicate two impulses of flare heating with each impulse characterized with an increase in the redshift. The emissions in the right panels of Figure 4 are warm-corona emissions that highlight the gradual phase of the flare and the corresponding response to the two separate heating periods during this flare. The Fe XIV (panel D) is most interesting in that it shows three peaks with what appears to be initial increase from the flare impulsive phase and then heated plasma about 10 minutes later after each of the transition-region redshifts peak for the two additional Fe XIV time series peaks. The Doppler shifts for those warm-corona emissions primarily are blue shifted with the two periods of maximum blueshifts corresponding to the two periods of maximum redshifts for the transition-region emissions.

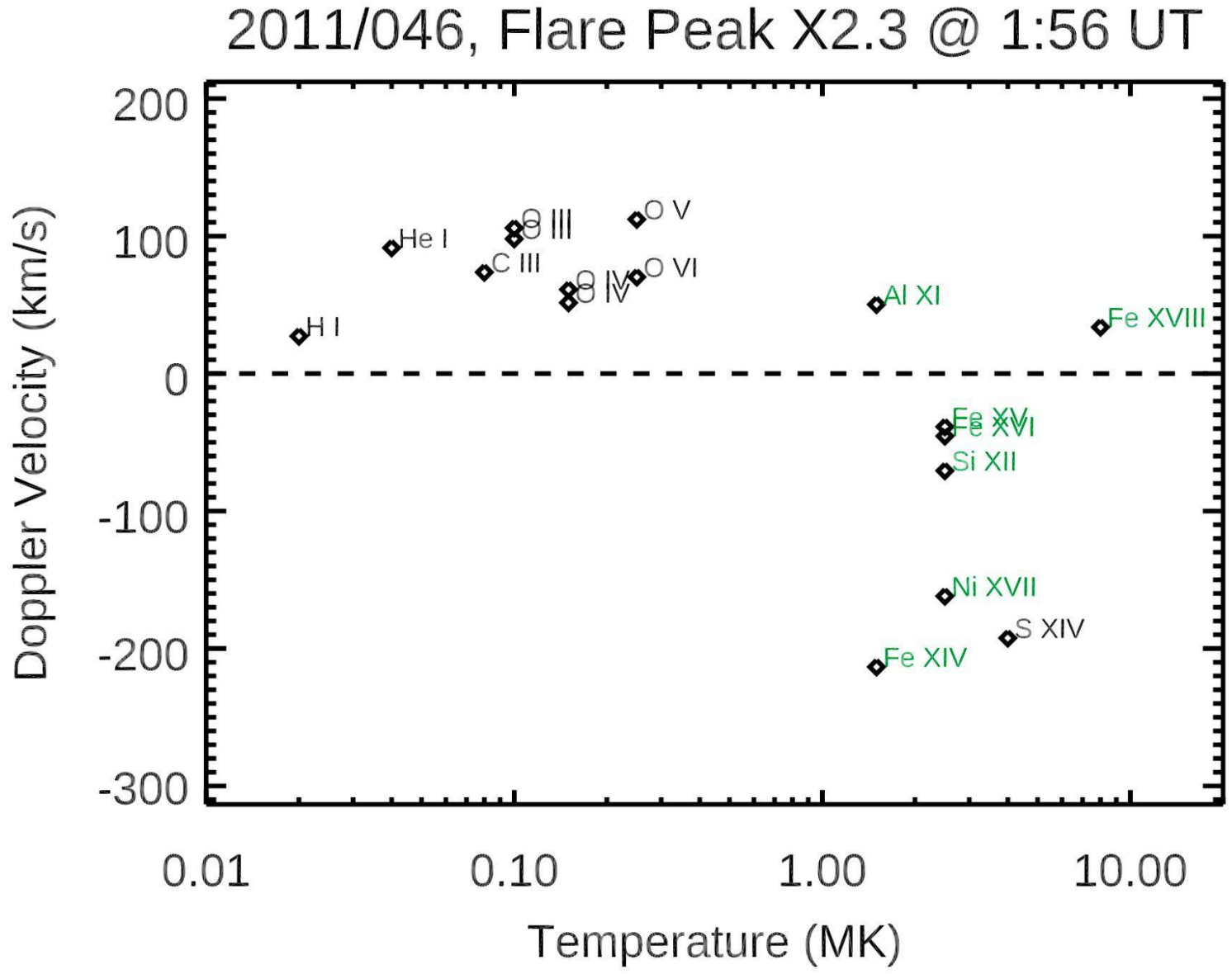

**Figure 3**. Maximum Doppler velocities during the X2.2 flare on 15 February 2011 as a function of line formation temperature. A positive Doppler velocity corresponds to a red wavelength shift, which signifies a downflow along the line of sight. A negative Doppler velocity corresponds to a blueshift (upflow).

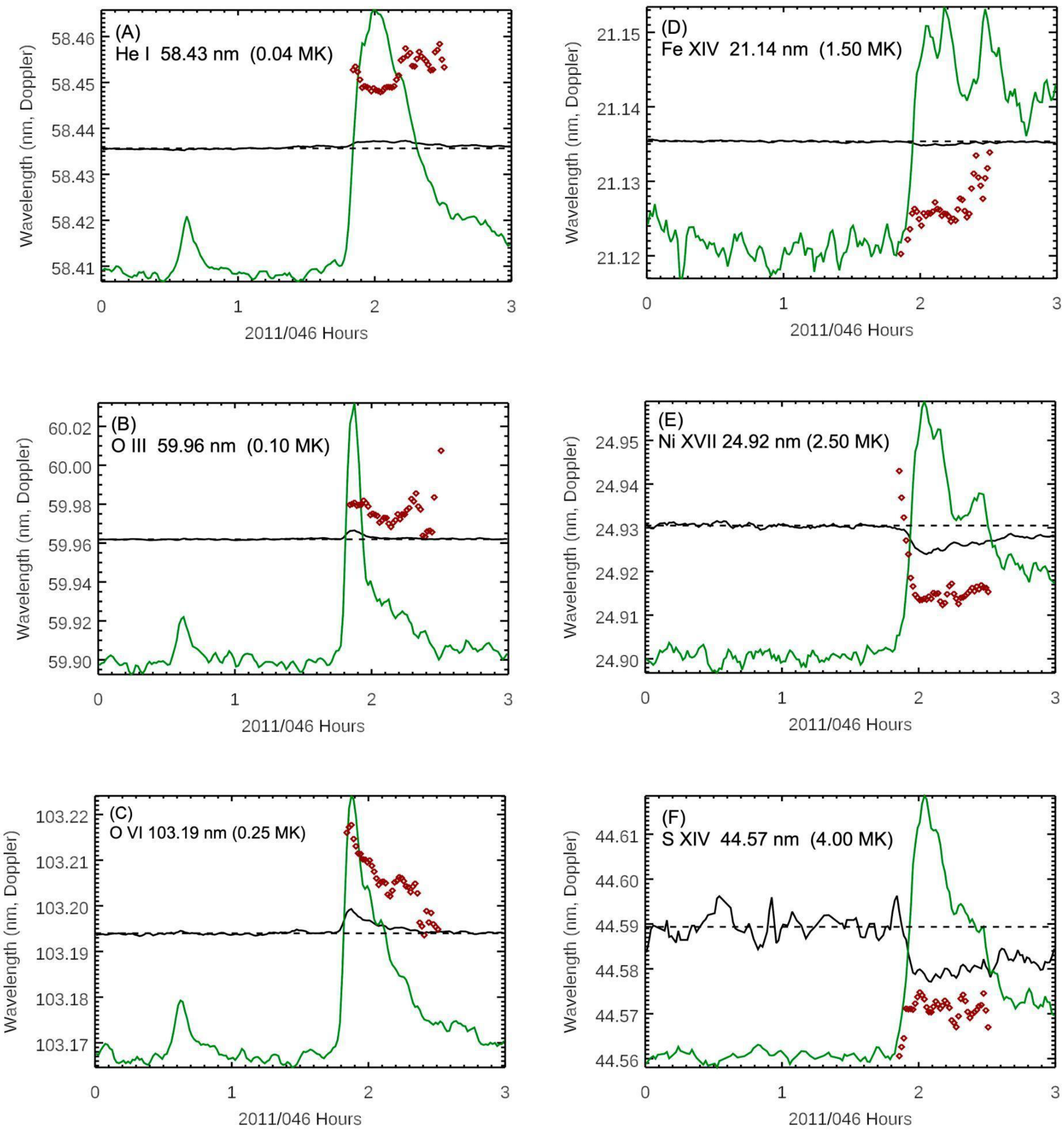

**Figure 4**. Doppler shifts during the X2.2 flare on 15 February 2011 for six emissions features in ascending formation temperature for (A) He I, (B) O III, (C) O VI, (D) Fe XIV, (E) Ni XVII, and (F) S XIV. The black lines are the central wavelengths fit to the full-disk irradiance, and the red diamonds are the central wavelengths fit to the flare irradiance (full-disk spectrum minus pre-flare spectrum). The dashed horizontal line is the pre-flare reference wavelength. The green lines are scaled irradiances for each emission feature.

In addition, these Doppler velocity results for this X2.2 flare are also consistent in magnitude of about 50 km/sec to other flare analyses with EVE Doppler velocity shifts presented by Hudson et al. (2011), Brown et al. (2016), and Otsu and Asai (2024). In particular, Hudson et al. (2011) shows a redshift of 50 km/sec for He II (30.4 nm) and a blue shift of about 100 km/sec for Fe XXIV (19.2 nm). Brown et al. (2016) examined the H Lyman lines and C III (53.8 nm) Doppler shifts and found shifts in range of 20-50 km/sec with redshifts for three flares and blueshifts for three different flares. As another example, Otsu and Asai (2024) studied the EVE O V (63.0 nm) and O VI (103.2 nm) features during the M8.7 flare on 2 October 2022 and found a large -400 km/sec blueshift during the filament eruption associated with this flare and then followed by a redshift during the flare decay phase as downflows for the cooling post-flare loops.

One possible application for these solar Doppler velocity shifts seen in the EVE spectra is the possible application to stellar studies, such as comparing solar and stellar flare dynamics. In particular, the ultraviolet emissions above the H continuum 91.1 nm cutoff are accessible for stellar observations. While most of those ultraviolet emission features are from the chromosphere and transition region, there are a few coronal emissions in the far ultraviolet range (e.g., Redfield et al., 2003) to consider for stellar Doppler velocity studies.

## 7. Coronal Dimming as Proxy for CME

The EUV late phase (see Section 4) is one way that solar eruptive events can feature extended periods of coronal evolution well after the main flare peak. Coronal dimming is another such phenomenon, commonly persisting for several hours following the initial energy release. Various physical processes can result in measurements characteristic of coronal dimming, but the one that is of most use for space weather is dimming that results from the plasma evacuation that is a coronal mass ejection (CME). Prior to the launch of SDO, it was not expected that there would be enough dimming to measurably reduce the total output of the sun at any particular wavelength (that is, the irradiance), but EVE has measured it.

In EUV and X-ray images, dimming looks much like the name it was originally given, transient coronal holes, with similar spatial extent but tending to last 3-12 hours (Reinard & Biesecker, 2008); whereas, traditional coronal holes persist for days-weeks. Dimming regions frequently map to the apparent base of CMEs as seen in white-light coronagraphs, providing a footprint of the eruptive process on the solar disk. In solar spectral irradiance, CME-induced dimming manifests as multiple emission lines near the ambient coronal temperature dipping nearly simultaneously. Hotter emission tends to come from confined bright loops, which have very characteristic flare light curves; that is, the hot corona emission doesn't show dimming nor is it expected to, as most of the mass of the CME is coming from higher up in the corona where the temperatures are closer to ambient.

Coronal dimming encodes information about the CME that caused it. Parameterizations of dimming can be related to the kinematics of the CME. For example, the spatial extent of the dimming is fairly well correlated with the mass of the CME (Dissauer et al., 2019) and in irradiance, the magnitude and rate of dimming are related to the CME mass and speed, respectively (Mason et al., 2014, 2016). In simple terms, the more mass a CME evacuates the deeper the drop in emission should be and the larger extent it is likely to have, and the faster a CME evacuates the steeper the decline in emission should be. The empirical results derived in the previously cited studies have also been supported by MHD modeling that simulates both the CME and the resultant dimming (Jin et al., 2022).

One challenge in spectral irradiance dimming measurements, like those made by EVE, is that dimming-sensitive emission lines are also sensitive to flare emission, so the light curves convolve these competing effects. Mason et al. (2014) demonstrated that not only can this effect be deconvolved given simultaneous multi-wavelength observations, but that we can disentangle true mass-loss dimming from confounding effects like temperature changes or obscuration. The clearest mass-loss dimming is seen as a near-simultaneous drop across multiple coronal emission lines, most notably Fe IX 17.1 nm and Fe XII 19.5 nm. By pairing these with flare-sensitive, dimming-insensitive lines (e.g., Fe XV 28.4 nm), it is possible to subtract out thermal contributions and isolate the dimming signal related to CME ejection, as shown in the example coronal dimming events in Figure 5. In a statistical follow-up, Mason et al. (2016) analyzed 37 dimming events and established quantitative relationships. These relationships establish the feasibility of estimating CME mass and speed in near real time from Sun-as-a-star irradiance observations, and in particular for events on the solar disk that would tend to be halo CMEs in coronagraphs on the same line-of-sight, which are both the hardest to obtain speeds (and thus arrival times) for and the most relevant for space weather since those are the events headed for Earth (if the observation is made along the Sun-Earth line). The CME mass and speed predictions from the example coronal dimming events shown in Figure 5 are listed in Table 2, and those results are in good agreement with CME properties determined with coronagraph observations.

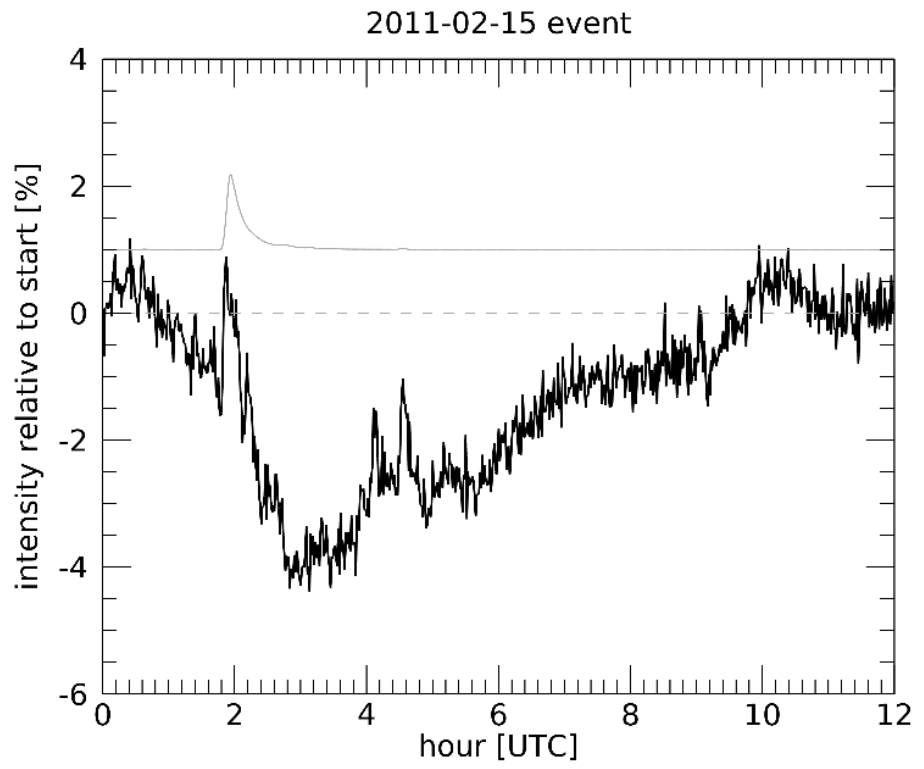

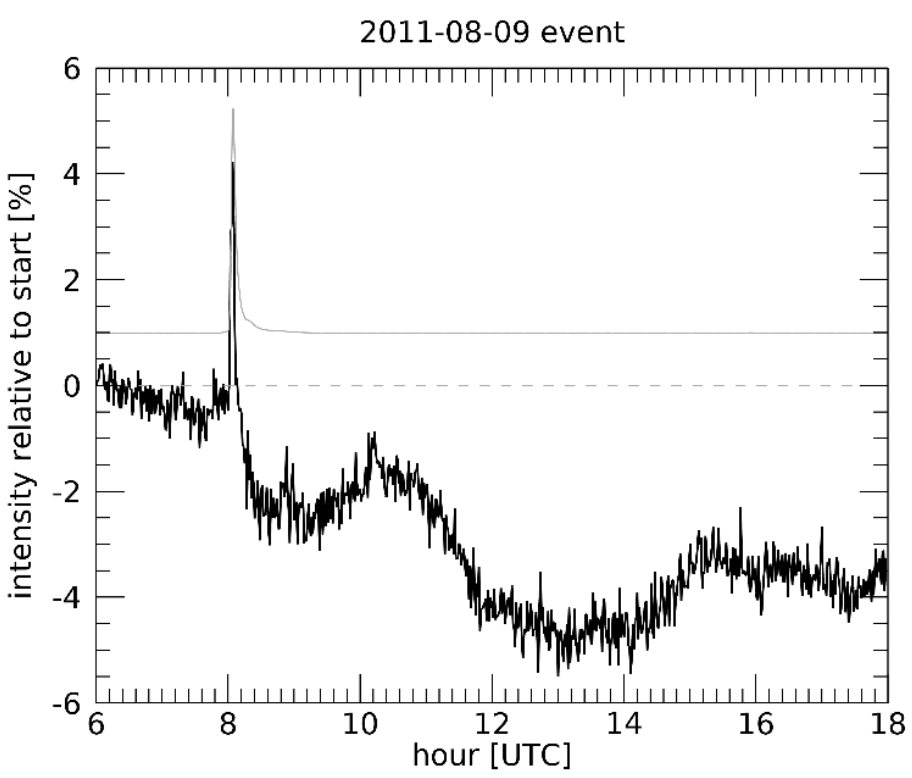

**Figure 5**. Examples of EVE coronal dimming measurements are shown for the X2.2 flare on 15 February 2011 (left) and for the X6.9 flare on 9 August 2011 (right). Both of these time series are the EVE measurements of the Fe IX 17.1 nm emission feature corrected for its impulsive phase and gradual phase contributions using the EVE Fe XV 28.4 nm emission (see Mason et al., 2014 for this correction). The light-grey line is the GOES XRS-B irradiance time series scaled so that it fits on the plot, and the dashed line is the pre-flare level. Note that the coronal dimming starts before the XRS irradiance significantly increases and that recovery takes several hours. There are additional decreases after the X6.9 flare (8:10 UT peak) due to additional flare / CME events at 9:12 UT and 11:26 UT.

**Table 2.** Example CME mass and speed predictions from EVE coronal dimming observations.

| Event Date | CME Mass Predicted from EVE Dimming | CME Mass Estimated from Coronagraph | CME Speed Predicted from EVE Dimming | CME Speed Estimated from Coronagraph |
|---|---|---|---|---|
| 15 Feb. 2011 | $5.4 \times 10^{15}$ g | $5.1 \times 10^{15}$ g | 1069 km/s | 897 km/s |
| 9 Aug. 2011 | $3.4 \times 10^{15}$ g | $3.8 \times 10^{15}$ g | 2294 km/s | 1474 km/s |

To date, statistical studies of the relationship between dimming and CMEs have had limited sample sizes (typically < 100 events). Mason et al. (2019) studied the entire history of SDO/EVE data during the era that MEGS-A was in operation (2010-2014), searching for and characterizing dimming that occurred around the time of more than 5000 ≥C1 flares. Subsequent studies to compare those dimmings with the associated CMEs are presently underway, funded by NASA's Astrophysics Decadal Survey Precursor Science program.

The detection of coronal dimming in full-Sun EUV irradiance has opened new opportunities for diagnosing CMEs on other stars—critical for understanding exoplanet habitability. While direct CME imaging is likely infeasible for stars, Veronig et al. (2021) showed that Sun-as-a-star SDO/EVE data can serve as a testbed for stellar CME detection. Their analysis of 44 large solar flares using EVE 15–25 nm integrated light curves demonstrated a strong statistical link: if a dimming is observed, the probability of an associated CME is ~97%. Most solar CMEs produced a measurable irradiance drop, and the rate of false detections (dimming without CME) was low, confirming dimming as a robust CME proxy.

Veronig et al. (2021) then extended this approach to stellar data from EUVE, XMM-Newton, and Chandra, applying the same analysis to a sample of active F, G, K, and M-type stars. They identified 21 dimming events on 13 stars, with depths up to 56% and durations of several hours—much stronger than typical solar events, mainly due to observational biases favoring large, clear signals. These stellar dimmings, occurring after flares, are strong candidates for stellar CMEs, suggesting that this method can provide valuable CME statistics even for unresolved stars.

This capability is directly relevant for exoplanetary science: frequent or massive stellar CMEs may drive atmospheric loss and affect the habitability of close-in planets. The synergy between solar EVE studies and stellar observations establishes coronal dimming as a practical tool for evaluating stellar space weather environments. As a result, a new astrophysics Small Explorer called ESCAPE (France et al., 2022) has been defined (but not selected for flight yet) to do for other stars what EVE has done for the Sun. The bandpass and spectral resolution for ESCAPE are very similar to EVE, and the telescope has been sized such that it can make high signal/noise measurements of even moderate dimming events occurring on other stars (Mason et al., 2025). EVE has made novel Sun-as-a-star measurements and that has inspired a new means of assessing the space weather around stars beyond the Sun.

## 8. Modeling EVE's Flare Variability

As noted in Sections 1 and 2, the SDO EVE observations were intended to be 24/7 with minimal amount of data gaps. While that has been the case for the EVE ESP channel, there have been unplanned issues with both the MEGS-A and MEGS-B channels, these 24/7 observations ended in the 5-37 nm MEGS-A band in 2014 and were limited to only ~3 hours/day in 36-105nm range with MEGS-B. In order to fill these temporal gaps in observational data, both for SDO and missions prior to SDO, solar VUV irradiance models are often used to provide complete temporal and spectral coverage. There are two main types of models used, physical and empirical models. Physical models are driven by fundamental plasma parameters such as plasma temperatures, emission measures (EM), densities, and abundances (see Section 5). Empirical models use a base set of measurements, when available, and build statistical relations and fits of these measurements to contemporaneously observed proxies. The proxies are a small set of measurements that are more often observed and for longer periods to cover the data gaps in the measurements. SDO EVE has become the critical reference dataset for developing empirical models of the X-ray and EUV irradiance because it is the most accurate and most complete solar EUV irradiance dataset. A handful of examples about modeling the solar EUV irradiance variability are discussed below.

Prior to SDO/EVE, the Flare Irradiance Spectral Model (FISM, Chamberlin et al., 2007, Chamberlin et al., 2008) was released in 2005 and used the predecessor to EVE, a mission called TIMED SEE (Woods et al., 2005), as its measurement baseline for the empirical model. The goal of FISM is to model the solar ultraviolet spectral irradiance over the full VUV range and over time scales from solar flares (seconds to hours), active region emergence (hours to days), solar rotation (days to weeks), and solar cycle (months to years). This original FISM was based only on 27 flares that SEE was able to observe with its 3% duty cycle, so the first version of FISM was less accurate for flare variability than its daily component.

Version 2 of the Flare Irradiance Spectral Model (FISM2, Chamberlin et al., 2020) was released in December 2020. The major improvements for this new version were provided by EVE's new measurements. These include 1) improved spectral resolution of EVE allowed the FISM spectral bins to improve by an order of magnitude from 1.0 nm to 0.1 nm, 2) more than 1,000 flares were observed at the time of FISM2 release by EVE to provide more accurate statistical fits for the empirical flare modeling, and 3) more accurate measurements from EVE inherently leads to more accurate FISM products with lower uncertainties, especially for the flare estimations.

Comparisons of the FISM2 predictions to the EVE Level 2 spectral measurements are shown in Figure 6 for the two example flares introduced in Section 4. Despite the significant improvement from the original FISM to FISM2, discrepancies remain between FISM2 and the measurements for which it is based. For this comparison, the FISM2 predictions slightly overestimate the flare irradiance for the X2.2 flare on 15 February 2011 (left panels) but more significantly underestimates the flare irradiance for the X6.9 flare on 9 August 2011 (right panels). One can also notice that FISM2 predictions for the hot corona emissions, like the Fe XX 13.3 nm emission (Figure 6 panels A and E), have flare peak times very similar to the EVE measurements, mainly because the FISM2 flare proxies are the GOES XRS irradiances that correlate very well with other hot coronal emissions. For cooler coronal emissions (Figure 6 panels B, C, F, and G), one will notice that the flare peaks later for these other EVE measurements than the FISM-2 predictions. These later peaks are expected for the warm coronal emissions while the post-flare loops cool. Because FISM2 flare proxies are only the GOES XRS irradiances, FISM2 does not predict the delayed peak times for those cooler emissions. We note that this issue for flare peaking time as a function of plasma temperature is being addressed for FISM3. Additionally, the coronal dimming and EUV late phase effects in the EVE solar EUV irradiance time series were not understood well when the FISM2 model was developed; consequently, those flare phases are not represented in the FISM2 algorithms.

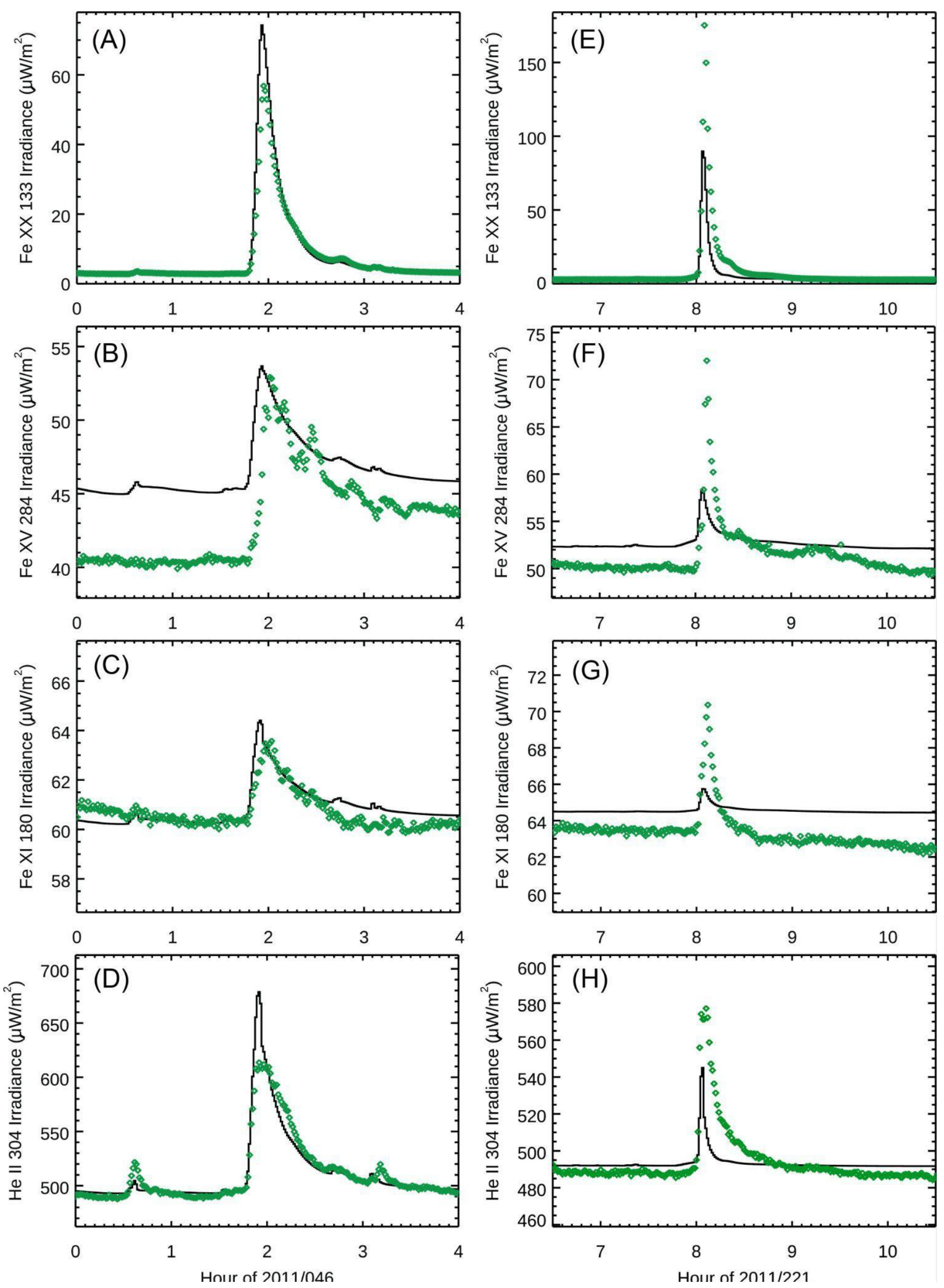

**Figure 6**. The FISM2 predictions (black lines) are compared to the EVE measurements (green diamonds) for the X2.2 flare on 15 February 2011 (left panels) and for the X6.9 flare on 9 August 2011 (right panels) for the four emission features highlighted in Figure 1 (solar flare phases). The emission features are integrated over 0.3nm-full-width band for this comparison.

FISM3, which is currently in development as a NASA Living With a Star (LWS) research grant, will address most of those differences noted in the comparison of FISM-2 predictions and EVE measurements (see Figure 6) and will overall improve the accuracy of predicting the solar VUV spectral irradiance. The specific improvements planned for FISM3 are (1) to incorporate the Lumped Element Thermal Model (LETM, Thiemann et al., 2017) for the delay in flare peak emissions for different wavelengths, (2) adding the additional daily proxies of plasma temperature and emissions measures (Schwab et al., 2023), (3) incorporate MinXSS CubeSats SXR spectral measurements (Woods et al., 2017, 2023), (4) incorporate the new routine measurements that are now available from the GOES EUVS instrument (Eparvier et al., 2009), (5) incorporate the ADAPT full-Sun magnetic field model (Hickmann et al., 2015), as well as relations of the photospheric magnetic field to irradiance, to drive a 'daily average' forecast of the VUV solar spectral irradiance.

Additionally, with the maturation of artificial intelligence (AI), deep learning algorithms are becoming more transparent (explainable AI) and are ready to address outstanding issues in Heliophysics, one of which is the development of FISM-AI. The results of FISM-AI will be directly compared to measurements, as well as to the improved FISM3 model to see which of the different methods provide the most accurate estimates for the solar VUV irradiance variability.

From a physics-based modeling approach, solar flares are widely understood to result from magnetic reconnection occurring over time, producing complex, multi-threaded flare structures. Numerical models have been developed to simulate the evolution of flare plasmas, representing flares as a series of impulsively heated loops (Warren, 2006). In these models that estimate the EUV variability during flares, the instantaneous heating rates and volumes are derived from GOES XRS light curves, while loop lengths are estimated using empirical footpoint separation models. Such modeling has successfully reproduced numerous aspects of high-temperature plasma evolution, including recent comparisons with soft X-ray spectra observed by the MinXSS CubeSat (Reep at al., 2020). However, the reliance on zero-dimensional hydrodynamic approximations restricts their ability to accurately capture dynamic processes within the solar chromosphere and transition region.

To address these limitations, Reep et al. (2022) implemented full one-dimensional hydrodynamic simulations using HYDRAD. Their simulations successfully reproduced the emission characteristics and temporal evolution of high-temperature coronal plasma ($T > 10$ MK). However, spectral lines formed at lower temperatures ($T < 1$ MK) in the chromosphere and transition region showed significant discrepancies with EVE observations, with simulated intensities exceeding observed values by 1-2 orders of magnitude. This discrepancy was most severe when assuming uniform cross-sectional loop areas. Reep et al. (2022) demonstrated that allowing cross-sectional area to vary with height along the magnetic field lines substantially improved agreement with observations. The remaining discrepancies likely arise from additional physical effects including non-uniform loop geometries, optical depth effects in strong spectral lines, or non-thermal electron distributions during the impulsive phase.

## 9. Flare Impacts in Earth's Ionosphere and Thermosphere

The ionosphere-thermosphere-mesosphere (ITM) system is an externally driven system, with solar irradiance mainly in the EUV and FUV as the primary external forcing (Knipp et al., 2005).

Solar EUV is absorbed in the thermosphere through photoionization of the major species N2, O, and O2, which creates the ionosphere, and solar EUV and FUV is absorbed through photodissociation of the molecular species. Energetic photoelectrons from photoionization can further ionize, dissociate, and excite neutral species. The products of these processes initiate a sequence of elastic and inelastic processes that transfer the initial photon energy into kinetic energy of the system. During solar flares, the rapid release of magnetic energy stored in the solar atmosphere causes a sudden (minutes) burst of solar SXR and EUV radiation from the Sun's surface and corona. This sudden enhancement of SXR and EUV takes about 8 minutes arriving at the Earth's upper atmosphere, which increases ionization and dissociation instantly, causing changes in composition, temperature, dynamics, and electrodynamics in the ITM system (e.g., Mitra, 1974; Zhang et al., 2002; Huba et al., 2005; Pawlowski & Ridley, 2011; Qian et al., 2011, 2012; Sojka et al., 2013; Le et al., 2015).

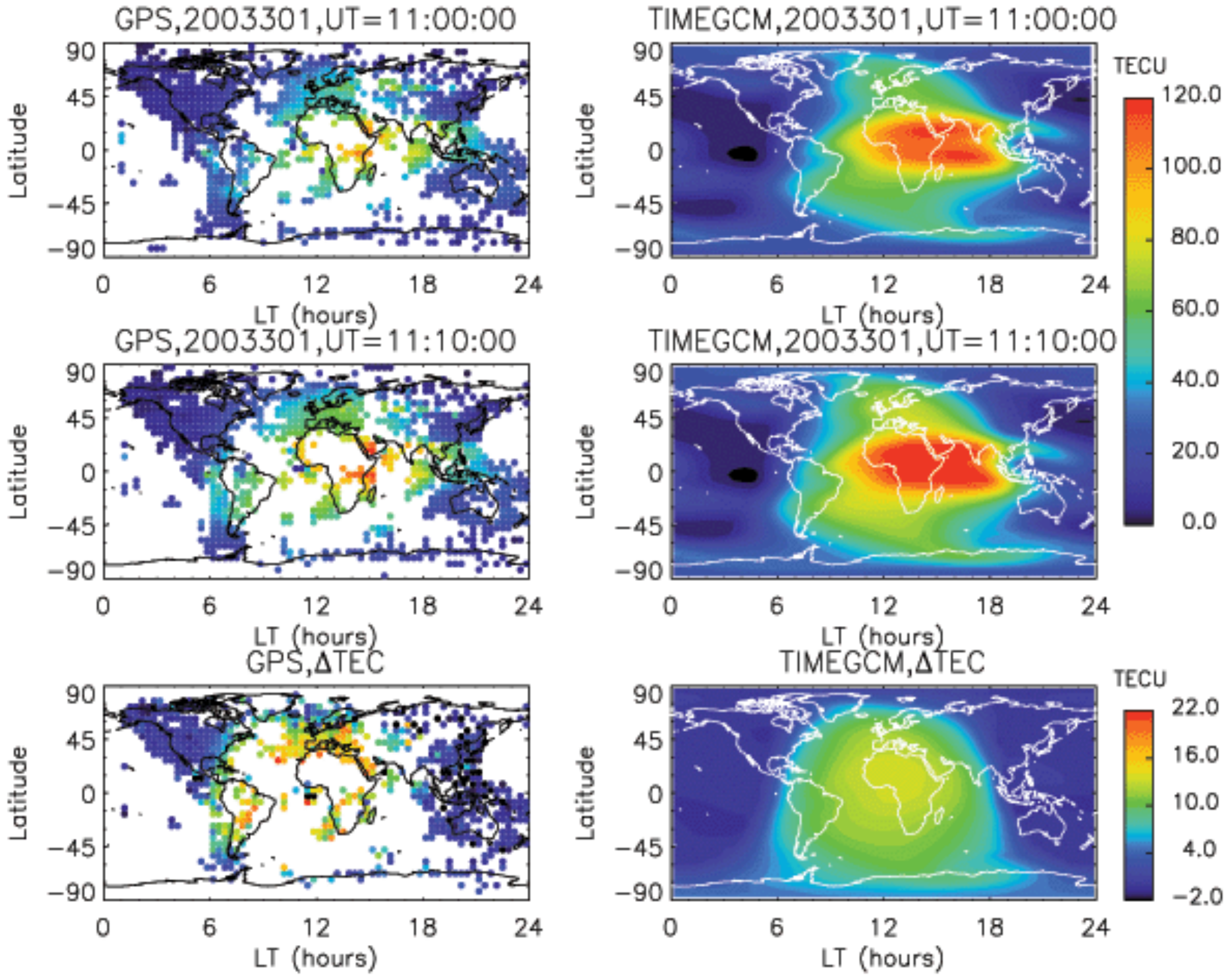

**Figure 7**. Comparisons of TEC observed by ground-based GPS network and TEC simulated by TIME-GCM in response to the X17 flare occurred on 28 October 2003. Left: GPS measurements of TEC for pre-flare, flare peak, and TEC enhancement; right: TIME-GCM simulations of TEC for pre-flare, flare peak, and TEC enhancement. Pre-flare: ∼11:00 UT; flare peak: ∼11:10 UT. DTEC equals the TEC difference between 11:10 and 11:00 UT on 28 October 2003 minus TEC difference between 11:10 and 11:00 UT on 27 October 2003. One TEC unit is $1x10^{12}$ electrons/cm$^2$. Adapted from Qian et al. (2010).

As an example of flare impacts in the ITM, observational and modeling studies revealed that in the ionosphere, the Global Positioning System (GPS) observed a 30% increase of total electron content (TEC) within 5 minutes in response to an X17 flare that occurred on 28 October 2003, and lasted for about 3 hours (e.g., Tsurutani et al., 2005). Figure 7 shows ~30% enhancement of TEC response to this X17 flare on 28 October 2023, along with ITM modeling comparison (Qian et al., 20110). This sudden increase in plasma density is a concern for space weather operations as it can impact, and even disrupt sometimes, GPS and High-Frequency (HF) radio signal propagation.

As another example, thermosphere mass density can increase ~ 10 – 60% in response to X-class flares at the altitudes of the low-Earth orbiting satellites. This mass density enhancement occurs instantly as flares occur, reaches the peak in 1 – 2 hours, and takes about 12 hours to recover (e.g., Liu et al., 2006; Sutton et al., 2006). These disturbances in the ITM system affect radio communication, GPS navigation, and satellite orbits (e.g., Mitra, 1974; Sutton et al., 2006). Figure 8 shows mass density enhancement of about 50% in response to the X17 flare on October 28, 2003 as observed by CHAMP and simulated by TIME-GCM. This sudden increase in neutral density in the low earth orbit environment causes sudden increase in satellite drag, which in turn speeds up satellites while lowering satellite orbit altitude.

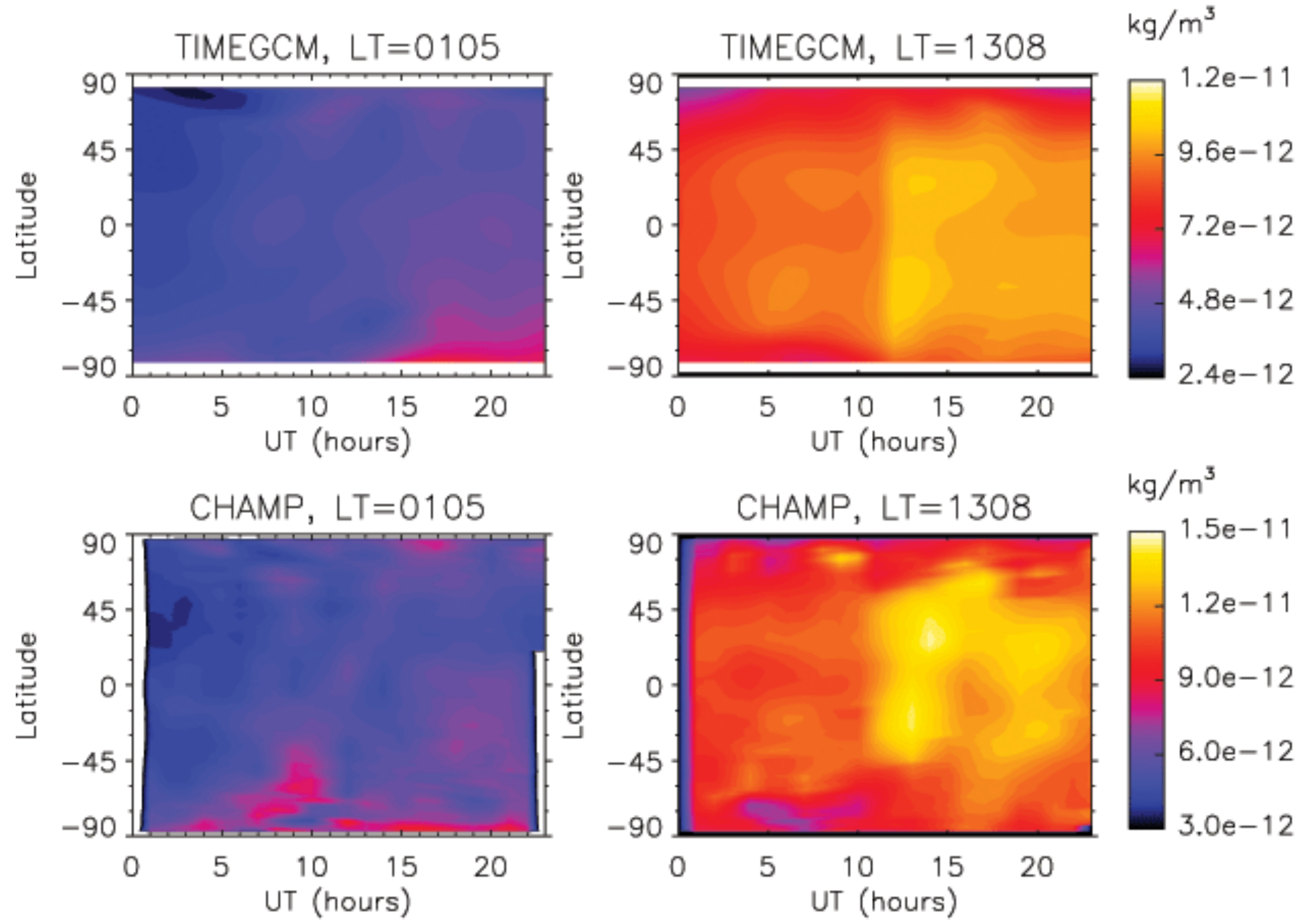

**Figure 8.** Comparisons of TIME-GCM simulated neutral density enhancement to CHAMP observed neutral density enhancement responding to the X17 flare occurred on 28 October 2003, during (left) CHAMP ascending node (0105 LT) and (right) CHAMP descending node (1308 LT). The flare started ~09:50 UT and peaked at ~11:10 UT. FISM solar spectra were used as solar input for the TIME-GCM. Adapted from Qian et al. (2010).

Solar flare impacts are not limited in the ITM system where they initially occur. The flare effects in the ITM system feed back into Earth's magnetosphere through the coupling between the ITM system and the magnetosphere. Figure 9 shows the model simulated magnetospheric convection in response to the X9.3 solar flare that occurred on 7 September 2017 (Figure 9a), the simulated magnetospheric convection where there is no flare (Figure 9b), and the difference (Figure 9c) in geocentric solar magnetospheric (GSM) coordinates. It is evident that the flare causes magnetospheric convection to be asymmetric with respect to noon-midnight meridian (the X-axis), with a significantly enhanced convection velocity in the pre-midnight plasma sheet compared to the post-midnight plasma sheet. This indicates a higher rate of reconnection along the pre-midnight x-line compared to the post-midnight x-line. This change in magnetospheric plasma convection changes energy flow from the magnetosphere to the ITM system through changes in both auroral particle precipitation and electromagnetic energy.

The flare effects in the magnetosphere is a feedback effect from the flare effects in the ITM system through the coupling between the ITM system and the magnetosphere. Liu et al. (2021) found during the X9.3 flare on 7 September 2017 that the simulated peak response of the pedersen conductance increased ~ 100%. This increased pedersen conductance increased the field aligned current by up to ~ 30%, reduced the cross polar cap electric potential by ~ 12.5%, and reduced Joule heating by ~ 50%. The reduced Joule heating indicates that the solar flare impacted energy dissipation in the ITM system because it reduced the effectiveness of the transfer of electromagnetic energy from solar wind/magnetosphere to thermal and kinetic energy in the ITM system.

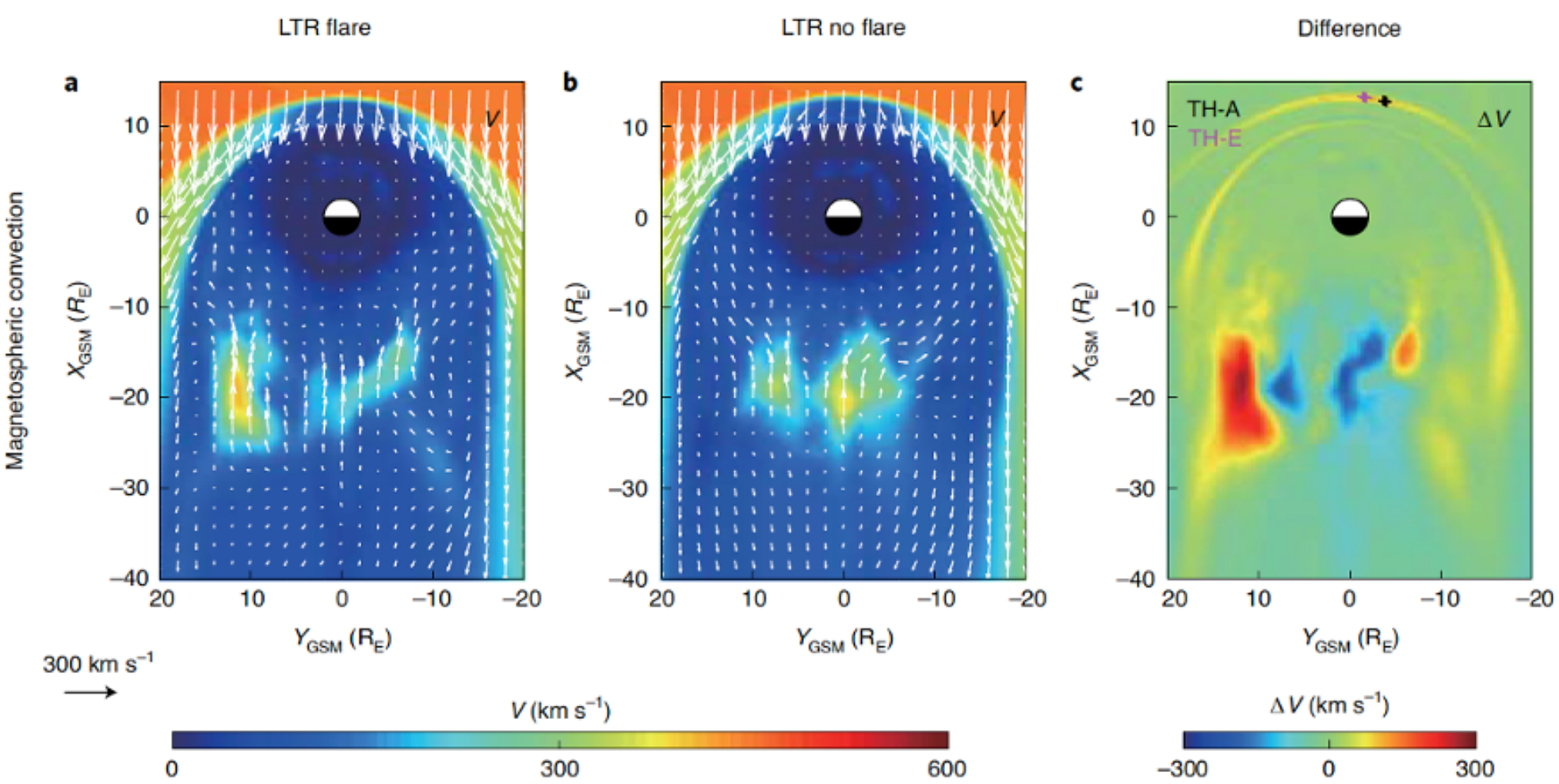

**Figure 9**. Solar flare effects on magnetospheric convection. Comparison of 50 min averages (12:02–12:51 UT) from model simulations of magnetospheric convection on 6 September 2017 with and without the X9.3 solar flare. Projections in geocentric solar magnetospheric (GSM) coordinates, with distances in Earth radii (RE), of model simulated magnetospheric convection velocity in the equatorial plane (ZGSM = 0) with (a) and without (b) solar flare effects and their difference (c). The black/white circle delineates the inner boundary of the simulation domain at a geocentric distance of 2 RE. The white half-circle represents the Sun-illuminated side. White arrows indicate direction and magnitude (also in color) of the convection velocity projected onto the plane. Adapted from Liu et al. (2021).

Previous studies found that the geoeffectiveness of solar flares depend on flare locations on the solar disk (e.g., Zhang et al., 2002; Qian et al., 2010). As an example, Figure 10 shows thermosphere mass density responses to the X17 flare and X28 flare that occurred on 28 October 2003 and 4 November 2003, respectively. The X17 flare was near the center of the solar disk (Figure 10a), whereas the X28 flare was on the limb (Figure 10b). Consequently, the thermospheric mass density enhancement during the X17 on-disk flare is significantly larger than the enhancement during the X28 limb flare.

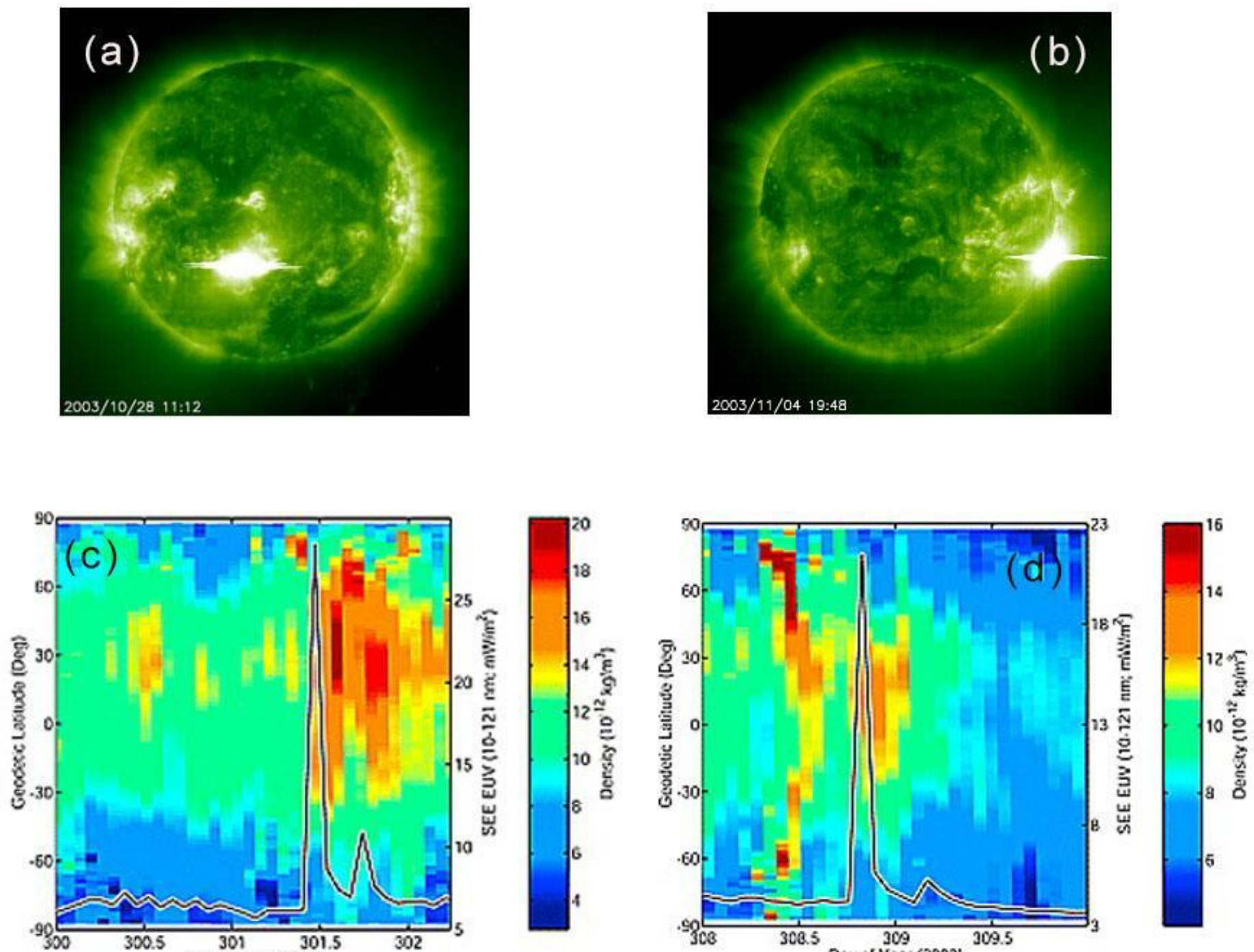

**Figure 10**. (a) SOHO EIT EUV image of the X17 flare that occurred on 28 October 2003; (b) SOHO EIT EUV image of the X28 flare that occurred on 4 November 2003; (c) mass density response to the X17 flare measured by CHAMP, normalized to 400 km, at local time 13:20; (d) mass density response to the X28 flare measured by CHAMP, normalized to 400 km. Solar EUV irradiances from the TIMED SEE instrument are superimposed in (c) and (d). The (c) and (d) panels are adapted from Sutton et al. (2006).

Other flare characteristics that impact flare geoeffectiveness include flare rise and decay times (e.g., Qian et al., 2011), and flare total energy (e.g., Pawlowski & Ridley, 2011). In addition, solar flares can occur preceding or during the geomagnetic storm main phases, as well as during the storm recovery phases. In such cases, the interplay between solar flares and geomagnetic storms causes additional complexity in how the ITM system responds to these events (e.g., Qian et al., 2012).

Furthermore, flares with EUV late phase (Woods et al., 2011; Woods, 2014) can also impact flare responses in the ITM system. Qian and Woods (2021) conducted model simulations to examine EUV late phase impacts in the thermosphere and ionosphere. They studied the responses to the C2.3, C8.8, and M1.3 flares on 5 May 2010. All three of those flares had an EUV late phase. To simulate how the flare EUV late phase affects the thermosphere and ionosphere responses, the flare variability is amplified by a factor of 10 for the model simulation. That is, the pre-flare irradiance level is the same, and just the flare variability (flare irradiance minus the pre-flare irradiance) is uniformly increased by a factor of 10 at each wavelength. This amplification increases the flare variability total energy by a factor of 10. These amplified flares are thus equivalent to M2.3, M8.8, and X1.3 flares. We refer to these flares as the big flares in Figure 11. The Figure 11a panel shows the integrated solar EUV irradiance (27 – 105 nm) on 5 May 2010 for the daily (black), the actual flares (blue), the big flares (red), and the big flares with the EUV late phases removed (cyan). Figure 11b is the solar heating increase due to the big flares, whereas Figure 11c shows the solar heating increase due to the big flares with the EUV late phases removed, at local noon, at 42.5°N. It is evident that the EUV late phase contributions have solar EUV fluctuations for the M2.3 and M8.8 flares, which then caused corresponding fluctuations in the solar heating. The extra EUV irradiance in the EUV late phase of the X1.3 flare produced a larger amount of extra heating above ~ 150 km where solar EUV radiation dominates the ionization.

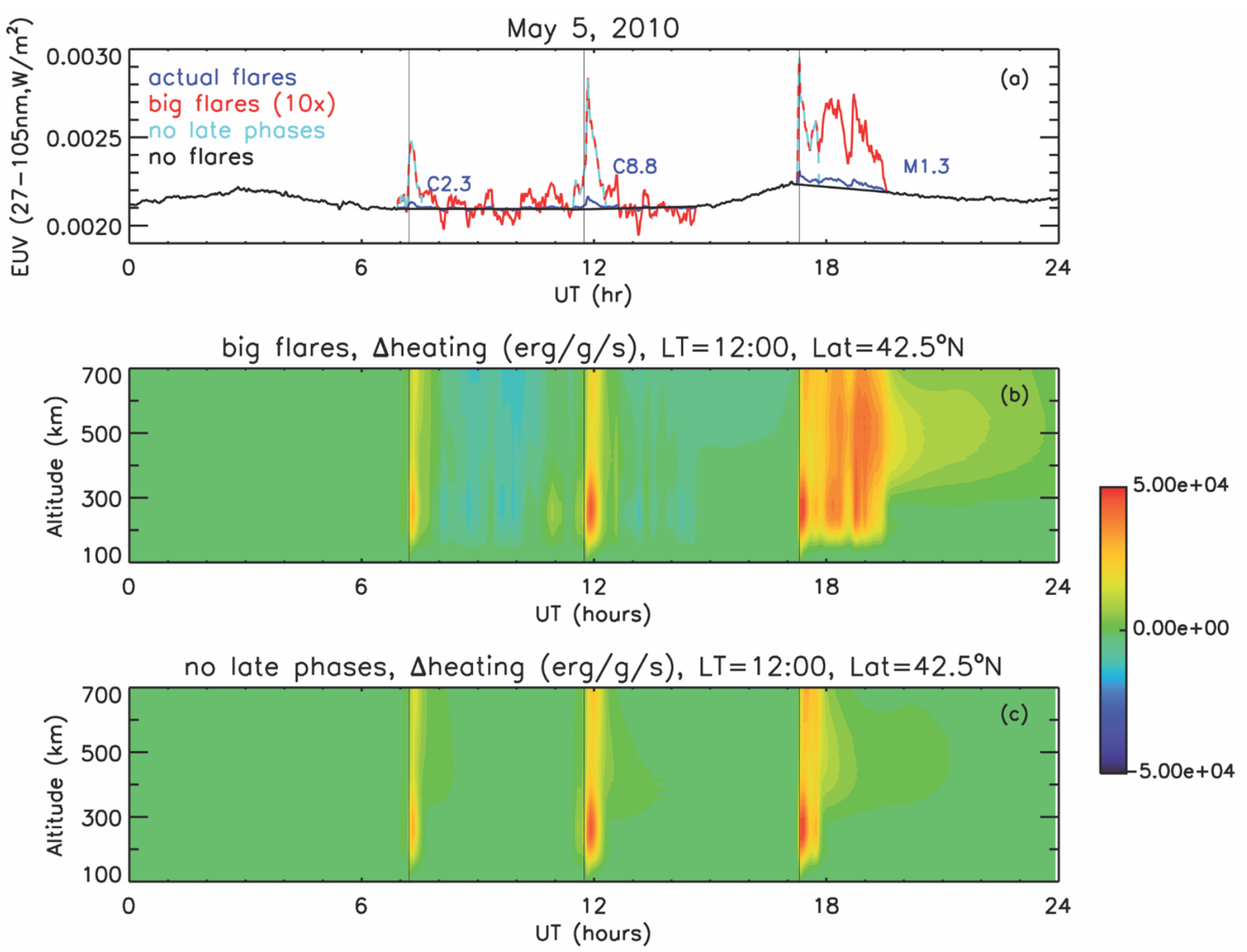

**Figure 11**. (a) EUV flux on 5 May 2010, provided by FISM. Blue: solar flux for the actual flares, C2.3, C8.8, and M1.3; black: daily solar flux, which does not include the flares; red: solar flux for the big flare, M2.3, M8.8, and X1.3; green: solar flux for the big flares with the EUV late phases removed; (b) TIME-GCM simulated thermosphere heating change in response to the big flares, at local noon and 42.5°N; (c) TIME-GCM simulated thermosphere heating change in response to the big flares with the EUV late phases removed, at local noon and 42.5°N. Adapted from Qian and Woods (2021).

## 10. Conclusions and Future Studies

The following are some key conclusions and future studies organized in order of the previous section topics.

There are five primary flare phases observed in the EVE solar EUV spectra as follows:

- Onset Phase: hot corona emissions indicate > 10 MK plasma is present at low emission-measure levels for several minutes before the main flare erupts,
- Impulsive Phase: eruptive flares usually have strong impulsive phase contributions from the transition region and cool corona emissions,
- Gradual Phase: most EUV and SXR emissions have a gradual phase contribution as the flaring loops are heated and then cool down and the cooler emissions have their peaks later than the hotter corona emissions,
- Coronal Dimming: eruptive flares associated with CMEs have dimming for the cool corona emissions from Fe XI through Fe XII and this dimming usually starts during the impulsive phase and can last for hours after the gradual phase, and
- EUV Late Phase: warm coronal emissions, such as from Fe XV and Fe XVI, can have a second peak by many minutes after the gradual phase peak, but without second peaks for hot corona emissions, and that more than 20% of M and X class flares can be characterized as EUV Late Phase flares.

There are several emission features that exemplify well the different flare phases, based mostly by the emission's plasma temperature and thus where the emission is formed within the solar atmosphere. The EVE MEGS-A "flare-phase" emission features were mostly highlighted here, but there are many other emission features in the EVE spectra and also from different instruments that can be used to study the various flare phases, especially in light that the MEGS-A observations ended in June 2014. Table 3 provides a list of some "flare-phase" emission features for other instruments as examples for which emissions could be used in future studies.

One important goal for future studies about flare phases is to better understand the flare onset phase and its potential for forecasting (nowcasting) flare magnitude several minutes into the future. Hudson et al. (2021) introduced the Hot Onset Precursor Event (HOPE) concept, which has been further confirmed by Battaglia et al. (2023), da Silva et al. (2023), and Telikicherla et al. (2024). However, the SDO observations do not address well the flare onset phase due to limitations of observing the hot corona above 10 MK, thus SXR and HXR observations from other instruments will be valuable for future flare onset phase studies. Currently in orbit, there are SXR observations from GOES XRS (Woods et al., 2025b), Hinode X-Ray Telescope (XRT) (Golub et al., 2007), Solar Orbiter Spectrometer/Telescope for Imaging X-rays (STIX, Krucker et al., 2020), Chandrayaan-2 Solar X-ray Monitor (SXM, Mithun et al., 2020), and Aditya-L1 Solar Low-Energy X-ray Spectrometer (SoLEXS, Sankarasubramanian et al., 2025). In the future, there will be new solar SXR measurements from the NASA CubeSat Imaging X-ray Solar Spectrometer (CubIXSS) CubeSat and NSF Impulsive Phase Rapid Energetic Solar Spectrometer (IMPRESS) CubeSat missions.

Because there is a factor of two to four in the difference for the photosphere abundance and coronal abundance for the low first ionization potential (low-FIP) elements (e.g., Fe, Si, Mg), studying the corona abundance changes during flares provide important diagnostics for the source location for coronal heating. Warren et al. (2014) clearly show that the abundances of the low-FIP elements, as derived from EVE spectra, are consistently close to photospheric values during flares. This important finding indicates that coronal heating, as part of the flare processes, is predominantly from evaporated plasma that originates deep within the chromosphere. This same conclusion has been confirmed from studying solar SXR spectra (e.g., Caspi et al., 2015; Telikicherla et al., 2024). It is promising that combining these multi-wavelength observations, as well as studying many more flare events, will lead to better understanding of the abundance changes, and thus coronal heating sources, throughout the full-lifecycle of flares as they have impulsive releases of energy into the corona and cool down in the post-flare loops. In addition, more studies are needed about the low-FIP abundance changes throughout the full-lifecycle of active region emergence and decay. For using the EVE full-disk irradiance measurements for active region studies, it is best to select times when only one active region dominates at a time, which is most feasible during solar minimum periods, such as in 2010 and 2019-2020 during the SDO mission.

**Table 3.** List of Emission Features that Exemplify the Different Flare Phases.

| Phase | Typical Temperature Range | EVE MEGS-A (ion) | EVE MEGS-B (ion) | EVE ESP Bands (ion) | SDO AIA Bands (ion) | GOES EXIS (channel) |
|---|---|---|---|---|---|---|
| Onset | 10-20 MK | 13.3 nm (Fe XX) | 54.1 nm (Fe XX) | 1-7 nm (SXR) | 13.1 nm (Fe XXI) | 0.1-0.8 nm (SXR) (XRS) |
| Impulsive | 0.1-0.5 MK | 30.4 nm (He II) | 53.7 nm (He I) | 30 nm (He II) | 30.4 nm (He II) | 30.4 nm (He II) (EUVS-A) |
| Gradual | 0.1-15 MK | 13.3 nm (Fe XX) | 54.1 nm (Fe XX) | 1-7 nm (SXR) | 13.1 nm (Fe XXI) | 0.1-0.8 nm (SXR) (XRS) |
| Dimming | 0.7-1.5 MK | 17.1 nm (Fe IX), corrected Fe XV 28.4 nm | 46.5 nm (Ne VII), 62.5 nm (Mg X) | 19 nm (Fe IX - Fe XIV) | 17 nm (Fe IX), 19 nm (Fe XII) | (none) |
| EUV Late | 2-4 MK | 33.5 nm (Fe XVI) | 50.0 nm (Si XII) | 28 nm (Fe XV) | 33.5 nm (Fe XVI) | 28.4 nm (Fe XV) |

In addition to studying the flare energetics (temperature, abundances), the EVE solar EUV spectra are being used to study the flare dynamics by measuring the Doppler shifts during flare events. Hudson et al. (2011) were the first to report that Doppler shifts are measurable with the EVE spectra, despite EVE not explicitly designed to measure Doppler shifts. The new EVE Level 4 Lines data product provides wavelength shift analysis for 70 of the emission lines in the EVE spectra with 60-s cadence, and 42 of those lines are considered good candidates for flare Dopper velocity studies (see Table 1). Corrections for MEGS instrumental wavelength shifts need to be applied to these EVE wavelength shift measurements if the flare is not near disk center. One can study the flare dynamics without making those MEGS instrumental corrections for flares near the disk center, as shown in Section 6. The flare events near solar disk center typically have red shifts (downflow) of about 50-100 km/s for the transition region emissions and blue shifts (upflows) of about 100-200 km/s for the warm corona emissions. These flare Doppler velocity shifts are largest during the flare's impulsive phase and are consistent with magnetic reconnection processes and chromospheric evaporation described by the standard CSHKP flare model.

It is interesting to note that a large prograde rotation Doppler shift was discovered in the EVE spectra by Hudson et al. (2022) for an active region rotating across the solar disk in 2012 and further analyzed by Fitzpatrick and Hudson (2023). They also find that the magnitude of the prograde rotation Doppler shift appears to increase with emission line formation temperature. We have recently discovered from new raytrace modeling of MEGS-B optical system with active regions at different positions across the solar disk that there are instrumental wavelength shifts (as much as +/- 35 km/s at some wavelengths) when the active regions or flares are near the limb. We are preparing an additional paper about the new EVE Level 4 Line Fit product and discussion about those MEGS-B raytrace model results. For now, flare dynamics studies with EVE spectra are most accurate for flares near disk center, and one should use caution (or at least larger uncertainties) for studies when the flares (or active regions) are near the limb.

Another surprise in the EVE spectra is that there is coronal dimming for the cooler corona (Fe IX through Fe XII) emissions during eruptive flares associated with CMEs. While corona dimming was known to occur over flares in solar EUV images, it was not expected to be detectable for full-disk irradiance measurements. As discussed in Section 7, the time series of the cooler corona features in the EVE spectra first need a correction (removal) of the emissions impulsive phase and gradual phase contributions by using a warmer corona (e.g. Fe XV) emission feature time series, and then the coronal dimming effects are clearly seen in the EVE observations (e.g., see Figure 5). The amount of the corona dimming is about 1-10% decrease in the irradiance for those cooler corona emission features. Two key results from studying the EVE coronal dimming measurements are that the coronal dimming magnitude is an indicator for CME mass and that the coronal dimming slope is an indicator for CME speed (Mason et al., 2016). These EVE-based CME mass and speed proxies are not considered as accurate as CME parameters derived with coronagraphs, but they offer a game-changing method to estimate CME properties on other stars as needed for studying stellar flares and habitability on exoplanets (e.g., Verinog et al., 2021). Additional studies to refine and improve the accuracy for coronal dimming results from the EVE solar spectra could lead to beneficial applications for space weather operations and also for future stellar EUV observations (e.g., France et al., 2025).

Two fundamental research modeling efforts planned for the EVE observations were to improve the solar variability models of the solar EUV and SXR spectral irradiance and to better understand through modeling the solar EUV and SXR impacts on Earth's ionosphere and thermosphere. As discussed in Section 8, the Flare Irradiance Spectral Model (FISM) was significantly improved in using the EVE solar EUV spectral irradiance measurements to make the FISM2 version. As shown in the FISM2 comparison in Figure 6, additional improvements to FISM are needed to address better the timing of the flare's gradual phase peaks as a function of emission temperature and to include the variability effects from coronal dimming and EUV late phase flare events. Those improvements are currently in development for inclusion into FISM3. Modeling of Earth's ionosphere and thermosphere response to solar variability is improved with the use of the FISM estimates because FISM can fill temporal gaps and spectral gaps over the full SXR, EUV, and FUV ranges.

As shown in Section 9, the large X-class flares can change the ionosphere plasma density by more than 20%, which in turn can directly impact GPS and HF radio signal propagation through the ionosphere. In addition, the large X-class flares can heat the thermosphere, which then expands upward to significantly change the thermosphere neutral density at satellite LEO altitudes. Knowing the satellite drag changes is a key factor for accurately tracking satellites as part of NOAA and Air Force space weather operations. Furthermore, modeling the flare effects in Earth's upper atmosphere have also revealed that solar flare effects in the ionosphere and thermosphere also couple into Earth's magnetosphere, that the solar flare location on the solar disk (near disk center versus on the limb) has different altitude impacts in Earth's atmosphere, and that EUV late phase flares had larger impacts above 150 km in Earth's atmosphere than flares without an EUV late phase. The EVE measurements and related improvements to solar EUV irradiance variability models, such as FISM2, have enabled these advances in understanding the solar EUV impacts in Earth's upper atmosphere. From similar solar-atmosphere modeling as presented in Section 9, Sojka et al. (2013) recommend new measurements of the solar SXR irradiance with modest spectral resolution for the 0.1-6 nm range that EVE MEGS spectra did not observe. With further improvements in understanding the solar EUV and SXR variability and thus improvements for the FISM3 estimates, future modeling of Earth's ionosphere and thermosphere are expected to further improve our understanding of the complex solar-atmosphere interactions, which in turn can further advance our forecast / nowcast capabilities and accuracies for space weather operations.

The SDO mission is long past its 5-year mission goal and has observed the Sun for over 15 years now. The SDO spacecraft and instruments (AIA, EVE, HMI) are all healthy, and so extending the SDO mission for many more years is feasible, but further extensions depend on NASA funding. The most recent NOAA GOES series of four satellites (GOES-16, GOES-17, GOES-18, and GOES-19) have similar solar EUV observations like SDO but have more limited wavelength coverage than SDO. Those NOAA GOES missions are expected to be operational through 2035 for the NOAA Space Weather Prediction Center. The GOES EXIS instrument (Eparvier et al., 2016; Woods et al., 2025a) has the XRS instrument for SXR irradiances in two bands (0.05-0.4 nm and 0.1-0.8 nm), the EUVS-A spectrograph for 25.6 nm, 28.4 nm, and 30.4 nm irradiances, the EUVS-B spectrograph for 117.5 nm, 121.6 nm, 133.5 nm, and 140.5 nm irradiances, and the EUVS-C spectrograph for the Mg II core-to-wing index (0.1 nm resolution from 279.55 nm to 280.27 nm). The GOES SUVI solar EUV imager (Darnel et al., 2022) obtains images at 9.4 nm, 13.1 nm, 17.1 nm, 19.5 nm, 28.4 nm, and 30.4 nm. Whereas, SDO AIA (Lemen et al., 2012) obtains solar EUV images at 9.4 nm, 13.1 nm, 17.1 nm, 19.3 nm, 21.1 nm, 30.4 nm, and 33.5 nm and also solar FUV images at 160 nm and 170 nm. NASA has plans to fly the Joint EUV coronal Diagnostic Investigation (JEDI) solar EUV imager on the ESA Vigil L5 mission in 2031, and NASA also plans to launch the Sun Coronal Ejection Tracker Concept CubeSat (SunCET, Mason et al., 2021) in 2026 with a wide-field solar EUV imager with a focus on studying CME acceleration with off-disk imaging of CMEs in 17-19 nm band. NOAA is considering next generation solar EUV imagers and spectrographs to replace the current GOES solar EUV instruments after 2035. It is not clear if there will be any opportunity over the next decade to fly an instrument with EVE-type capability, so continuation of the SDO mission is important to continue its unique monitoring of the solar EUV spectral irradiance. With EVE's capability to detect the different flare phases, plasma temperature, emission measure, low-FIP elemental abundances, flare dynamics through measuring Doppler shifts, and CME parameters using coronal dimming of cool-corona emission features, there is still much to study with EVE spectra and other instruments to further advance our understanding about solar flare physics.

## Acknowledgements

The SDO EVE original development, mission operations, data processing, and science research have been supported by NASA contract NAS5-02140 to the University of Colorado. H. P. Warren was supported by the Office of Naval Research, and J. P. Mason was partially supported by NASA Grant 80NSSC25K7012. The authors gratefully acknowledge the many people who have contributed to the success of the EVE instrument throughout development and mission operations and to operating the SDO mission for more than 15 years. We are also very thankful for the extensive international community who have expanded the knowledge of solar physics through analysis and modeling of the SDO data sets.

## Data Sources

The primary data shown in this manuscript are SDO EVE solar EUV irradiance measurements that are available as public EVE data products (version 8.1) at http://lasp.colorado.edu/eve/ . The GOES XRS data are from https://www.ncei.noaa.gov/products/goes-r-extreme-ultraviolet-xray-irradiance/ , and SDO AIA solar EUV images are from http://jsoc.stanford.edu/AIA/AIA_jsoc.html and also https://suntoday.lmsal.com/suntoday/ . The CHIANTI spectral model is available from https://www.chiantidatabase.org/chianti_download.html . The TIME-GCM model is available from https://www.hao.ucar.edu/modeling/tgcm/ .

## References

Acton, L. W., 2016, On-Orbit performance and calibration of the Soft X-Ray Telescope on Yohkoh, *Solar Phys.*, 291, 643. https://doi.org/10.1007/s11207-015-0842-5

Aschwanden, M. J., Boerner, P., Caspi, A., McTiernan, J. M., Ryan, D., Warren, H., 2015, Benchmark test of differential emission measure codes and multi-thermal energies in solar active regions, *Solar Phys*., 290, 2733. https://doi.org/10.1007/s11207-015-0790-0

Battaglia, A. F., Hudson, H., Warmuth, A., et al., 2023, The existence of hot X-ray onsets in solar flares, *Astron. Astrophys*., 679, id.A139. https://doi.org/10.1051/0004-6361/202347706

Benz, A. O., 2017, Flare observations, *Living Rev. Sol. Phys.*, 14, id 2. https://doi.org/10.1007/s41116-016-0004-3

Berretti, M., Mestici, S., Giovannelli, L., et al., 2025, ASR: Archival Solar Flares catalog, *Astrophys. J. Suppl*., 278, id.9. https://doi.org/10.3847/1538-4365/adc731

Brosius, J. W., Phillips, K. J. H., 2004, Extreme-ultraviolet and X-ray spectroscopy of a solar flare loop observed at high time resolution: A case study in chromospheric evaporation, *Astrophys. J*., 613, 580. https://doi.org/10.1086/422873

Brown, S. A., Fletcher, L., Labrosse, N., 2016, Doppler speeds of the hydrogen Lyman lines in solar flares from EVE, *Astron. Astrophys*., 596, id.A51. https://doi.org/10.1051/0004-6361/201628390

Caspi, A., McTiernan, J. M., Warren, H. P., 2014, Constraining solar flare differential emission measures with EVE and RHESSI, *Astrophys. J. Lett*., 788, id.L31. https://doi.org/10.1088/2041-8205/788/2/L31

Chamberlin, P. C., 2016, Measuring solar Doppler velocities in the He II 30.38 nm emission using the EUV Variability Experiment (EVE), *Solar Phys*., 291. 1665. https://doi.org/10.1007/s11207-016-0931-0

Chamberlin, P. C., Woods, T. N., Eparvier, F. G., 2006, Flare Irradiance Spectral Model (FISM): Daily component algorithms and results, *Space Weather,* 5, S07005. https://doi.org/10.1029/2007SW000316

Chamberlin, P. C., Woods, T. N., Eparvier, F. G., 2008, Flare Irradiance Spectral Model (FISM): Flare component algorithms and results, *Space Weather,* 6, S05001. https://doi.org/10.1029/2007SW000372

Chamberlin, P. C., Milligan, R. O., Woods, T. N., 2012, Thermal evolution and radiative output of solar flares observed by the EUV Variability Experiment (EVE), *Solar Phys*., 279, 23. https://doi.org/10.1007/s11207-012-9975-y

Chamberlin, P. C., Eparvier, F. G., Knoer, V., et al., 2020, The Flare Irradiance Spectral Model - Version 2 (FISM2), *Space Weather*, 18, 12. https://doi.org/10.1029/2020SW002588

Culhane, J. L., Hiei, E., Doschek, G. A., et al., 1991, The Bragg Crystal Spectrometer for SOLAR-A, *Solar Phys.*, 136, 89. https://doi.org/10.1007/BF00151696

Czaykowska, A., De Pontieu, B., Alexander, D., Rank, G., 1999, Evidence for chromospheric evaporation in the late gradual flare phase from SOHO/CDS observations, *Astrophys. J*., 521, L75. https://doi.org/10.1086/312176

Del Zanna, G., Woods, T. N., 2013, Spectral diagnostics with the SDO EVE flare lines, *Astron. Astrophys*., 555, id.A59. https://doi.org/10.1051/0004-6361/201220988

Del Zanna, G., Dere, K.P., Young, P.R., Landi, E., Mason, H.E., 2015, CHIANTI – an atomic database for emission lines. Version 8, *Astron. Astrophys.*, 582, A56. doi.org/10.1051/0004-6361/201526827

Dere, K.P., Landi, E., Mason, H.E., Monsignori Fossi, B.C., Young, P.R., 1997, CHIANTI – an atomic database for emission lines, *Astron. Astrophys. Suppl.*, 125, 149. doi.org/10.1051/aas:1997368

Darnel, J. M., Seaton, D. B., Bethge, C., et al., 2022, The GOES-R Solar UltraViolet Imager, *Space Weather*, 20, e2022SW003044. https://doi.org/10.1029/2022SW003044

Didkovsky, L. V., Judge, D. L., Wieman, S., Woods, T., Jones, A., 2012, The Extreme Ultraviolet SpectroPhotometer (ESP) in Extreme Ultraviolet Variability Experiment (EVE): Algorithms and calibrations, *Solar Phys.*, 275, 179. https://doi.org/10.1007/s11207-009-9485-8

Dissauer, K., Veronig, A. M., Temmer, M., Podladchikova, T., 2019, Statistics of coronal dimmings associated with coronal mass ejections. II. Relationship between coronal dimmings and their associated CMEs, *Astrophys. J.*, 874, 123. https://doi.org/10.3847/1538-4357/ab0962

Donnelly, R. F., 1987, Temporal trends of solar EUV and UV full-disk fluxes, *Solar Phys.*, 109, 37. https://doi.org/10.1007/BF00167398

Eparvier, F. G., Crotser, D., Jones, A. R., et al., 2009, The Extreme Ultraviolet Sensor (EUVS) for GOES-R, *Proc. SPIE: Solar Physics and Space Weather Instrumentation III*, 7438, 743804. https://doi.org/10.1117/12.826445

Fitzpatrick, J. C., Hudson, H. S., 2022, The temperature dependence of hot prograde flows in solar active regions, *Solar Phys.*, 298, id.2. https://doi.org/10.1007/s11207-022-02093-3

Fletcher, L., Dennis, B. R., Hudson, H. S., et al., 2011, An observational overview of solar flares, *Space Sci. Rev.*, 159, id.19. https://doi.org/10.1007/s11214-010-9701-8

Fludra, A., Schmelz, J. T., 1999, The absolute coronal abundances of sulfur, calcium, and iron from Yohkoh-BCS flare spectra, *Astron. Astrophys.*, 348, 286.

France, K., Fleming, B., Youngblood, A., et al., 2022, Extreme-ultraviolet stellar characterization for atmospheric physics and evolution mission: Motivation and overview, *J. Astron. Tel. Instr. Sys.*, 8, 014006. https://doi.org/10.1117/1.JATIS.8.1.014006

Garcia, H. A., 2000, Temperature and emission measure from GOES soft X-ray measurements, *Solar Phys.*, 193, 33. https://doi.org/10.1023/A:1005213923962

Golub, L., DeLuca, E., Austin, G., et al., 2007, The X-ray Telescope (XRT) for the Hinode mission, *Solar Phys.*, 243, 63. https://doi.org/10.1007/s11207-007-0182-1

Gonzalez, G., Chamberlin, P., Herde, V., 2024, *Solar Phys*., 299, 151. https://doi.org/10.1007/s11207-024-02394-9

Gopalswamy, N., Akiyama, S., Yashiro, S., Mäkelä, P., 2010, Solar flares with and without coronal mass ejections, *Astrophys. J.*, 710, 1111. https://doi.org/10.1088/0004-637X/710/2/1111

Handy, B. N., Acton, L. W., Kankelborg, C. C., et al. , 1999, The Transition Region and Coronal Explorer — mission overview and first results, *Solar Phys.*, 187, 229. https://doi.org/10.1023/A:1005166902804

Hanser, F. A., Sellers, F. B., 1996, Design and calibration of the GOES-8 Solar X-ray Sensor: The XRS, *Proc. SPIE: EUV, X-Ray, and Gamma-Ray Instrumentation for Astronomy VII,* 2812, 344. https://doi.org/10.1117/12.254077

Hickmann, K. S., Godinez, H. C., Henney, C. J., Arge, C. N., 2015, Data assimilation in the ADAPT photospheric flux transport model, *Solar Phys.*, 290, 1105. https://doi.org/10.1007/s11207-015-0666-3

Hock, R. A., 2012, The Role of Solar Flares in the Variability of the Extreme Ultraviolet Solar Spectral Irradiance, PhD Dissertation, Univ. Colorado at Boulder, Boulder, CO. https://www.proquest.com/pqdtglobal/docview/1017865491/fulltextPDF/138CEA9CA8984B84PQ/1?accountid=14503

Hock, R. A., Chamberlin, P. C., Woods, T. N., et al., 2012, Extreme Ultraviolet Variability Experiment (EVE) MEGS-A radiometric calibration, *Solar Phys.,* 275, 145. https://doi.org/10.1007/s11207-010-9520-9

Hock, R. A., Woodraska, D., Woods, T. N., 2013, Using SDO EVE data as a proxy for GOES XRS-B 1-8 Angstrom, Space Weather, 11, 262. https://doi.org/10.1002/swe.20042

Huba, J. D., Joyce, G., Fedder, J. A., 2005, SAMI3: A Three-dimensional low-latitude ionosphere model, *J. Geophysical Res.,* 110, A12302. https://doi.org/10.1029/2005JA011213

Hudson, H. S., 2011, Global properties of solar flares, *Space Sci. Rev.*, 158, 5. https://doi.org/10.1007/s11214-010-9721-4

Hudson, H. S., 2025, Anticipating solar flares, *Solar Phys*., 300, id.2. https://doi.org/10.1007/s11207-024-02418-4

Hudson, H. S., Woods, T. N., Chamberlin, P. C., et al., 2011, The EVE Doppler sensitivity and flare observations, *Solar Phys.*, 273, 69. https://doi.org/10.1007/s11207-011-9862-y

Hudson, H. S., Simões, P. J. A., Fletcher, L., Hayes, L. A., Hannah, I. G., 2021, Hot X-ray onsets of solar flares, *Mon. Not. Royal Astron. Soc.*, 501, 1273. https://doi.org/10.1093/mnras/staa3664

Hudson, H. S., Mulay, S. M., Fletcher, L., et al., 2022, Fast prograde coronal flows in solar active regions, *Mon. Not. Royal Astron. Soc*., 515, L84. https://doi.org/10.1093/mnrasl/slac079

Jin, M., Cheung, M. C. M., DeRosa, M. L., Nitta, N. V., Schrijver, C. J., 2022, Coronal mass ejections and dimmings: A comparative study using MHD simulations and SDO observations, *Astrophys. J*., 928, 154. https://doi.org/10.3847/1538-4357/ac589b

Judge, D. L., McMullin, D., Ogawa, H. S., et al., 1998, First solar EUV irradiances obtained from SOHO by the CELIAS/SEM, *Solar Phys*., 177, 161. https://doi.org/10.1023/A:1004929011427

Klimchuk, J. A., Patsourakos, S., Cargill, P. J., 2008, Highly efficient modeling of dynamic coronal loops, *Ap. J.*, 682, 1351.. https://doi.org/10.1086/589426

Kniezewski, K. L., Mason, E. I., Uritsky, V. M., Garland, S. H., 2024, 131 and 304 Å emission variability increases hours prior to solar flare onset, *Astrophys. J. Lett.*, 977, id.L29. https://doi.org/10.3847/2041-8213/ad94dd

Knipp, D. J., Welliver, T., McHarg, M. G., et al., 2005, Climatology of extreme upper atmospheric heating events, Adv. Space Res., 36, 2506. https://doi.org/10.1016/j.asr.2004.02.019

Krucker, S., Battaglia, M., Jenke, P., et al., 2020, The Spectrometer/Telescope for Imaging X-rays (STIX) on Solar Orbiter, *Astron. Astrophys.*, 642, A15. https://doi.org/10.1051/0004-6361/202037362

Laming, J. M., 2004, A unified picture of the first ionization potential and inverse first ionization potential effects, *Astrophys. J.*, 614, 1063. https://doi.org/10.1086/423728

Landi, E., Young, P.R., Dere, K.P., Del Zanna, G., Mason, H.E., 2013, CHIANTI—an atomic database for emission lines. XIII. Soft X-ray improvements and other changes, *Astrophys. J.*, 763, 86. https://doi.org/10.1088/0004-637X/763/2/86

Le, H. J., Ren, Z., Liu, L., Chen, Y., Zhang, H., 2015, Global thermospheric disturbances induced by a solar flare: a modeling study, *Earth, Planets, Space*, 67, A3. https://doi.org/10.1186/s40623-014-0166-y

Lean, J. L., Woods, T. N., Eparvier, F. G., et al., 2011, Solar extreme ultraviolet irradiance: Present, past, and future, *J. Geophy. Res. Space Phys.*, 116, A01102. https://doi.org/10.1029/2010JA015901

Lemen, J. R., Title, A. M., Akin, D. J., et al., 2012, The Atmospheric Imaging Assembly (AIA) on the Solar Dynamics Observatory, *Solar Phys.*, 275, 17. https://doi.org/10.1007/s11207-011-9776-8

Liu, L., Wan, W., Maruyama, T., Zhao, B., Huang, F. J., 2006, Solar activity variations of the ionospheric peak electron density, *J. Geophys. Res.*, 111, A05303. https://doi.org/10.1029/2006JA011598

Liu, J., Wang, W., Qian, L., et al., 2021, Solar flare effects in the Earth's magnetosphere, Nat. Phys., 17, 807. https://doi.org/10.1038/s41567-021-01203-5

Mason, J. P., Woods, T. N., Caspi, A., Thompson, B. J., Hock, R. A., 2014, Mechanisms and observations of coronal dimming for the 2010 August 7 event, *Astrophys. J.*, 789, id.61. https://doi.org/10.1088/0004-637X/789/1/61

Mason, J. P., Woods, T. N., Webb, D. F., et al., 2016, Relationship of EUV irradiance coronal dimming slope and depth to coronal mass ejection speed and mass, *Astrophys. J.*, 830, 20. https://doi.org/10.3847/0004-637X/830/1/20

Mason, J. P., Attie, R., Arge, C. N., Thompson, B., Woods, T. N., 2019, The SDO/EVE solar irradiance coronal dimming index catalog. I. Methods and algorithms, *Astrophys. J. Suppl.*, 244, 13. https://doi.org/10.3847/1538-4365/ab380e

Mason, J. P., Chamberlin, P. C., Seaton, D. , et al., 2021, SunCET: The Sun Coronal Ejection Tracker Concept, *J. Sp. Weather Sp. Clim.*, 11, 20. https://doi.org/10.1051/swsc/2021004

Mason, J. P., Youngblood, A., France, K., Veronig, A., Jin, M., 2025, Detecting stellar coronal mass ejections via coronal dimming in the extreme ultraviolet, *Ap. J.*, 998, id.167. https://doi.org/10.3847/1538-4357/ade4bc

Milligan, R. O., 2014, Extreme ultraviolet spectroscopy of the lower solar atmosphere during solar flares, *Solar Phys.*, 290, 3399. https://doi.org/10.1007/s11207-015-0748-2

Milligan, R. O., Dennis, B. R., 2009, Velocity characteristics of evaporated plasma using Hinode/EUV Imaging Spectrometer, *Astrophys. J.*, 699, 968. https://doi.org/10.1088/0004-637X/699/2/968

Milligan, R. O., Kennedy, M. B., Mathioudakis, M., Keenan, F. P., 2012, Time-dependent density diagnostics of solar flare plasmas using SDO/EVE, *Astrophy. J. Lett.*, 755, L16. https://doi.org/10.1088/2041-8205/755/1/L16

Mithun, N. P. S., Vadawale, S. V., Sarkar, A., et al., 2020, Solar X-ray Monitor on board the Chandrayaan-2 Orbiter: In-flight performance and science prospects, *Solar Phys.*, 295, 139. https://doi.org/10.1007/s11207-020-01712-1

Mitra, A. P., 1974, *Ionospheric Effects of Solar Flares*, D. Reidel Publishing Company, Dordrecht/Boston.

Neupert, W. M., 1968, Comparison of solar X-ray line emission with microwave emission during flares, *Astrophys. J. Lett.*, 153, L59. https://doi.org/10.1086/180220

Otsu, T., Asai, A., 2024, Multiwavelength Sun-as-a-star analysis of the M8.7 Flare on 2022 October 2 using Hα and EUV spectra taken by SMART/SDDI and SDO/EVE, *Astrophys. J.*, 964, id.75. https://doi.org/10.3847/1538-4357/ad24ec

Pawlowski, D. J., Ridley, A. J., 2011, The effects of different solar flare characteristics on the global thermosphere, *J. Atmos. Sol. Terr. Phys.*, 73, 1840. https://doi.org/10.1016/j.jastp.2011.04.004

Petkaki, P., Del Zanna, G., Mason, H. E., Bradshaw, S. J., 2012, SDO AIA and EVE observations and modelling of solar flare loops, *Astron. Astrophys.*, 547, id.A25. https://doi.org/10.1051/0004-6361/201219812

Priest, E. R., Forbes, T. G., 2002, The magnetic nature of solar flares, *Astron. Astrophy. Rev.*, 10, 313. https://doi.org/10.1007/s001590100013

Qian, L., Woods, T. N., 2021, Solar flare effects on the thermosphere and ionosphere, in *Space Physics and Aeronomy Volume 4: Upper Atmosphere Dynamics and Energetics*, ed. W. Wang, Y. Zhang, L. Paxton, Geophys. Mono. Ser., 261, AGU Wiley, p. 253. https://doi.org/10.1002/9781119815631.ch14

Qian, L., Burns, A. G., Chamberlain, P. C., Solomon, S. C., 2010, Flare location on the solar disk: Modeling the thermosphere and ionosphere response, *J. Geophys. Res*., 115, A09311. https://doi.org/10.1029/2009JA015225

Qian, L., Burns, A. G., Chamberlain, P. C., Solomon, S. C., 2011, Variability of thermosphere and ionosphere responses to solar flares, *J. Geophys. Res*., 116, A10309. https://doi.org/10.1029/2011JA016777

Qian, L., Burns, A. G., Solomon, S. C., Chamberlain, P. C., 2012, Solar flare impacts on ionospheric electrodynamics, *Geophys. Res. Lett.*, 39, L06101. https://doi.org/10.1029/2012GL051102

Redfield, S., Ayres, T. R., Linsky, J. L., et al., 2003, A far ultraviolet spectroscopic explorer survey of coronal forbidden lines in late-type stars, *Astrophys. J.*, 585, 993. https://doi.org/10.1086/346129

Reep, J. W., Warren, H. P., Moore, C. S., Suarez, C., Hayes, L. A., 2020, Simulating solar flare irradiance with multithreaded models of flare arcades, *Astrophys. J*., 895, 30. https://doi.org/10.3847/1538-4357/ab89a0

Reep, J., Ugarte-Urra, I., Warren, H., Barnes, W. T., 2022, Geometric Assumptions in hydrodynamic modeling of coronal and flaring loops, *Astrophys. J.*, 933, 106. https://doi.org/10.3847/1538-4357/ac7398

Reinard, A. A., Biesecker, D. A., 2008, Coronal mass ejection - associated coronal dimmings, *Astrophys. J*., 674, 576. https://doi.org/10.1086/525269

Sankarasubramanian, K., Bug, M., Sarwade, A., et al., 2025, Solar Low-Energy X-ray Spectrometer (SoLEXS) on board Aditya-L1 mission, *Solar Phys.*, 300, 87. https://doi.org/10.1007/s11207-025-02494-0

Schrijver, C. J., Siscoe, G. L., editors, 2010, *Heliophysics: Space Storms and Radiation: Causes and Effects*, Cambridge Univ. Press, London, ISBN: 9780521760515.

Schwab, B. D., Woods, T. N., Mason, J. P., 2023, Modeling the daily variations of the coronal X-ray spectral irradiance with two temperatures and two emission measures, *Astrophys. J.*, 945, id.31, https://doi.org/10.3847/1538-4357/acb774

Shibata, K., Magara, T., 2011, Solar flares: Magnetohydrodynamic processes, *Living Rev. Sol. Phys.*, 8, id 6. https://doi.org/10.12942/lrsp-2011-6

Solomon, S. C., Qian, L., Woods, T. N., 2013, Solar extreme-ultraviolet irradiance for general circulation models, J. Geophysical Res., 118, 5881. https://doi.org/10.1002/jgra.50500

Sutton, E. K., Forbes, J. M., Nerem, R. S., Woods, T. N., 2006, Neutral density response to the solar flares of October and November 2003, *Geophys. Res. Lett.*, 33, L22101. https://doi.org/10.1029/2006GL027737

Telikicherla, A., Woods, T. N., Schwab, B. D., 2024, Investigating the soft X-Ray spectra of solar flare onsets, *Astrophys. J.*, 966, 198. https://doi.org/10.3847/1538-4357/ad37f6

Thiemann, E. M. B., Eparvier, F. G., Woods, T. N., 2017, A time dependent relation between EUV solar flare light-curves from lines with differing formation temperatures, *J. Sp. Weather Sp. Clim*., 7, A36. https://doi.org/10.1051/swsc/2017037

Thompson, B. J., Plunkett, S. P., Gurman, J. B., Newmark, S. S., St. Cyr, O. C., Michels, D. J., 1998, SOHO/EIT observations of an Earth-directed coronal mass ejection on May 12, 1997, *Geophys. Res. Lett*., 25, 2465. https://doi.org/10.1019/98GL50429

Tsuneta, S., Acton, L. W., Bruner, M. E., et al., 1991, The Soft X-ray Telescope for the SOLAR-A mission, *Solar Phys*., 136, 37. https://doi.org/10.1007/BF00151694

Tsurutani, B. T., Judge, D. L., Guarnier, F. L., et al., 2005, The October 28, 2003 extreme EUV solar flare and resultant extreme ionospheric effects: Comparison to other Halloween events and the Bastille Day event, *Geophys. Res. Lett*., 32, L03S09. https://doi.org/10.1029/2004GL021475

Veronig, A. M., Odert, P., Leitzinger, M., et al., 2021, Indications of stellar coronal mass ejections through coronal dimmings, *Nature Astron.*, 5, 697. https://doi.org/10.1038/s41550-021-01345-9

Warren, H. P., 2006, Multithread hydrodynamic modeling of a solar flare, *Astrophys. J.*, 637, 522. https://doi.org/10.1086/497904

Warren, H. P., 2014, Measurements of absolute abundances in solar flares, *Astrophys. J. Lett.,* 786, id.L2. https://doi.org/10.1088/2041-8205/786/1/L2

Warren, H. P., Mariska, J. T., Doschek, G. A., 2013, Observations of Thermal Flare Plasma with the EUV Variability Experiment, *Astrophys. J*., 770, id.116. https://doi.org/10.1088/0004-637X/770/2/116

Warren H. P., Brooks, D. H., Doschek, G. A., Feldman, U., 2016, Transition region abundance measurements during impulsive heating events, *Astrophys. J*., 824, id.56. https://doi.org/10.3847/0004-637X/824/1/56

White, S. M., Thomas, R. J., Schwartz, R. A., 2005, Updated expressions for determining temperatures and emission measures from GOES soft X-ray measurements, *Solar Phys.*, 227, 231. https://doi.org/10.1007/s11207-005-2445-z

Woods, T. N., 2014, Extreme ultraviolet late-phase flares: Before and during the Solar Dynamics Observatory mission, *Solar Phys.*, 289, 3793. https://doi.org/10.1007/s11207-014-0483-0

Woods, T. N., Elliott, J., 2022, Solar Radiation and Climate Experiment (SORCE) X-Ray Photometer System (XPS): Final data-processing algorithms, *Solar Phys*., 297, id.64. https://doi.org/10.1007/s11207-022-01997-4

Woods, T. N., Eparvier, F. G., Bailey, S. M., et al., 2005, Solar EUV Experiment (SEE): Mission overview and first results, *J. Geophysical Res.*, 110, A01312. https://doi.org/10.1029/2004JA010765

Woods, T.N., Chamberlin, P.C., Peterson, W.K., et al., 2008, XUV photometer system (XPS): improved solar irradiance algorithm using CHIANTI spectral models, *Solar Phys.*, 250, 235. https://doi.org/10.1007/s11207-008-9196-6

Woods, T. N., Hock, R., Eparvier, F., et al., 2011, New solar extreme-ultraviolet irradiance observations during flares, *Astrophys. J*., 739, 59. https://doi.org/10.1088/0004-637X/739/2/59

Woods, T. N., Eparvier, F. G., Hock, R., et al., 2012, Extreme Ultraviolet Variability Experiment (EVE) on the Solar Dynamics Observatory (SDO): Overview of science objectives, instrument design, data products, and model developments, *Solar Phys.*, 275, 115. https://doi.org/10.1007/s11207-009-9487-6

Woods, T. N., Caspi, A., Chamberlin, P. C., et al., 2017, New solar irradiance measurements from the Miniature X-ray Solar Spectrometer CubeSat, *Astrophys. J.*, 835, 122. https://doi.org/10.3847/1538-4357/835/2/122

Woods, T. N., Schwab, B., Sewell, R., et al., 2023, First results for solar soft X-Ray irradiance measurements from the third-generation Miniature X-Ray Solar Spectrometer, *Astrophys. J.*, 956, id.94. https://doi.org/10.3847/1538-4357/acef13

Woods, T. N., Eden, T., Eparvier, F. G., et al., 2024, The GOES-R Series X-Ray Sensor (XRS): Design and pre-flight calibration results, *J. Geophys. Res*., 129, 2024JA032925. https://doi.org/10.1029/2024JA032925

Woods, T. N., Woodraska, D., Eparvier, F. G., et al., 2025, *EVE Calibration and Measurement Algorithms Document*, Univ. of Colorado, Boulder. https://doi.org/10.25810/nh0m-ze62

Zhang, D. H., Xiao, Z., Igarashi, K., Ma, G. Y., 2002, GPS-derived ionospheric total electron content response to a solar flare that occurred on 14 July 2000, *Radio Sci*., 37, 1806. https://doi.org/10.1029/2001RS002542